\documentclass[a4paper,12pt,onecolumn,fleqn,twoside]{article}

\usepackage[headheight=0.6cm,left=2.5cm,right=1.5cm,top=2.0cm,bottom=2.0cm]{geometry}

\usepackage{times}

\usepackage[small,bf]{caption}
\usepackage{graphicx}

\usepackage{array}
\usepackage{dcolumn}
\newcolumntype{d}{D{.}{.}}

\usepackage[square,numbers,sort&compress]{natbib}

\usepackage{latexsym}
\usepackage{amsmath}
\usepackage{amssymb}
\usepackage{bbm}

\allowdisplaybreaks[1] 

   \newcommand{\expvalue}[1]{\ensuremath{\left \langle \Psi_0 \left | #1 \right | \Psi_0 \right \rangle}}
   \newcommand{\expvalueZero}[1]{\ensuremath{\left \langle 0 \left | #1 \right | 0 \right \rangle}}
   \newcommand{\expvaluePsi}[1]{\ensuremath{\left \langle \Psi \left | #1 \right | \Psi \right \rangle}}
   \newcommand{\expvalueNucleon}[1]{\ensuremath{\left \langle N \left | #1 \right | N \right \rangle}}
   \newcommand{\timeorder}[1]{\ensuremath{T \left [ #1 \right ]}}
   \newcommand{\condensate}[1]{\ensuremath{\left \langle #1 \right \rangle}}
   \newcommand{\vaccondensate}[1]{\ensuremath{\left \langle #1 \right \rangle_{\rm vac}}}
   \renewcommand{\slash}[1]{\ensuremath{ #1 \mspace{-8mu} / }}
   \newcommand{\lslash}[1]{\ensuremath{ #1 \mspace{-10mu} / }}
   \newcommand{\trace}[1]{\ensuremath{\mathrm{Tr} \left ( #1 \right ) }}
   \newcommand{\vackappa}[1]{\ensuremath{\kappa^{\rm vac}_{ \rm #1 }}}
   \newcommand{\medkappa}[1]{\ensuremath{\kappa^{\rm med}_{ \rm #1 }}}
   \newcommand{\tvackappa}[1]{\ensuremath{\tilde{\kappa}^{\rm vac}_{ \rm #1 }}}
   \newcommand{\tmedkappa}[1]{\ensuremath{\tilde{\kappa}^{\rm med}_{ \rm #1 }}}
   \newcommand{\tkappa}[1]{\ensuremath{\tilde{\kappa}}}
   \newcommand{\unit}[1]{\ensuremath{\; {\rm #1}}}
   \newcommand{\borel}{\ensuremath{ \mathcal{M} }}
   \newcommand{\indexrm}[1]{\ensuremath{ \mathrm{#1} }}
   \newcommand{\state}[1]{\ensuremath{\left | #1 \right \rangle}}

\newcommand{\condensateAlphaPiVGsquareplusVGtsquare}{\condensate{\dfrac{\alpha_s}{\pi} [ (vG)^2 + (v\tilde{G})^2 ] }}

\begin{document}
\title{Four-Quark Condensates in Nucleon QCD Sum Rules}

\author{R. Thomas$^{a}$, T. Hilger$^{b}$, B. K\"ampfer$^{a,b}$ \\
\small $^a$ Forschungszentrum Dresden-Rossendorf, PF 510119, 01314 Dresden, Germany \vspace{-5pt} \\
\small $^b$ Institut f\"ur Theoretische Physik, TU Dresden, 01062 Dresden, Germany \vspace{-5pt}
}

\date{}

\maketitle

\renewcommand{\abstractname}{}
\begin{abstract}
The in-medium behavior of the nucleon spectral density including self-energies is revisited within the framework of QCD sum rules. Special emphasis is given to the density dependence of four-quark condensates. A complete catalog of four-quark condensates is presented and relations among them are derived. Generic differences of such four-quark condensates occurring in QCD sum rules for light baryons and light vector mesons are discussed.
\end{abstract}

\section{Introduction}
A goal of contemporary hadron physics is to relate the confined quark and gluon degrees of freedom and parameters related to Quantum Chromodynamics (QCD) to the comprehensive hadronic spectrum. Lattice QCD and chiral effective field theory are suitable tools to accomplish this and other goals in exploring the structure of low-energy QCD and properties of hadrons. Another - though not so direct - but successful approach is given by QCD sum rules, originally formulated by Shifman, Vainshtein and Zakharov \cite{Shifman:1978bx} to describe masses of light vector mesons for example \cite{Shifman:1978by}. The method since then gained attention in numerous applications, e.\ g.\ to calculate masses and couplings of low-lying hadrons, magnetic moments, etc.\ (cf.\ e.g.~\cite{Reinders:1984sr,pk:Narison2004,Colangelo:2000dp}). Its particular meaning is that numerous hadronic observables are directly linked to a set of fundamental QCD quantities, the condensates and moments of parton distributions.

Hadrons are excitations from the ground state. Changes in this state are expected to reflect in a change of hadronic properties, especially in spectral functions and moments thereof related, e.g., to masses of hadrons.
Measurements of ''mass modifications'' of hadrons in a finite temperature, strongly interacting medium or when situated inside nuclear matter, that
means embedded in a bulk of protons and neutrons and baryonic and mesonic resonances, then probe the QCD vacuum (cf.~\cite{Krusche:2006ki} for an experimental overview). The properties of the QCD ground state are mapped to and quantified by a number of condensates, like intrinsic material constants, which partially carry information on symmetry features of the theory. Besides the chiral condensate $\condensate{\bar{q} q}$, up to mass dimension 6 the gluon condensate $\condensate{\tfrac{\alpha_s}{\pi} G^2 }$, the mixed quark-gluon condensate $\condensate{\bar{q} g_s \sigma G q}$, the triple gluon condensate $\condensate{g_s^3 G^3}$ and structures of the form $\condensate{\bar{q} \Gamma q \bar{q} \Gamma q}$ contribute in vacuum. ($\Gamma$ denotes all possible structures formed by Dirac and Gell-Mann matrices.) We emphasize here the specific role of the latter class of hitherto poorly known condensates, the four-quark condensates.

Within the realm of hadron spectroscopy the explanation of the actual numerical value of the nucleon mass is crucial as ingredient for understanding macroscopic matter. The nucleon represents as carrier of mass the hard core of visible matter in the universe and thus is an important source for gravitation.
Our investigations here are to be considered in line with previous investigations~\cite{Furnstahl:1992pi,Jin:1992id,Jin:1993up,Cohen:1994wm} for nucleons inside cold nuclear matter, which are also discussed in~\cite{Henley:1993nr} and continuously explored in~\cite{Drukarev:1988kd,Drukarev:1994fw,Drukarev:2001wd,Drukarev:2003xa,Drukarev:2004zg,Drukarev:2004fn,Sadovnikova:2005ye,Sadovnikova:2006te}.
The possible extension to further effects at finite temperatures lies beyond the present scope.
An advantage of applying QCD sum rules is that effects of small finite baryon density $n$ are described systematically by the change of condensates and the advent of new structures which are absent at zero density.
For the nucleon an important dependence of self-energies on four-quark condensates was found. Comparisons with results of chiral effective field theory~\cite{Gross-Boelting:1998jg}, where nucleon self-energies show strong cancellation effects (i.\ e.\ they change with the same magnitude but have opposite signs) suggest that the relevant four-quark condensates should be weakly density dependent~\cite{Furnstahl:1992pi}.
For the $\omega$ meson, however, we recently deduced in~\cite{Thomas:2005dc} evidence for a significant density dependence of a particular combination of four-quark condensates appearing there. This was based on the experimentally found shift of the spectral strength of $\omega$ to lower invariant masses inside nuclear matter, as measured by the CB-TAPS collaboration~\cite{Trnka:2005ey}.
Therefore, we will spell out explicitly the four-quark condensates in nucleon sum rules in medium, which up to now are usually given in the factorization approximation or are determined by special models. So one can directly distinguish the four-quark condensate structures in sum rules for vector mesons and baryons.

The work is organized as follows. In section 2 we review the operator product expansion for the nucleon and the hadronic model to write out the QCD sum rule equations. The four-quark condensates are discussed in section 3 where an exhaustive list is presented and relations among these condensates are given. Afterwards we present a numerical analysis of their influences in a nucleon sum rule evaluation and compare to other results (section 4). Conclusions can be found in section 5. In the appendices, an explanation of the calculation of an OPE and remarks on four-quark condensate relations are supplemented.

\section{QCD Sum Rules for the Nucleon}
\label{qcdsumrulesforthenucleon}
\subsection{Current-Current Correlator}
QCD sum rules (QSR) link hadronic observables and expectation values of quark and gluon operators. This allows to determine properties of the low-lying hadronic excitations of the QCD ground state $\state{\Psi}$.
It relies on the concept of semi-local quark-hadron duality applied to the time-ordered correlation function
\begin{equation}
\Pi (q) = i \int d^4x \, e^{iqx} \expvaluePsi{\timeorder{ \eta (x) \bar{\eta} (0) }} \equiv i \int d^4x \, e^{iqx} \Pi (x) ,
\label{eq:correlationFunction}
\end{equation}
which describes the propagation of a hadron created from the vacuum. On one side it can be calculated in terms
of quarks and gluons via an operator product expansion (OPE) for large space-like momenta $q^2$. This introduces Wilson coefficients multiplied by local normal ordered expectation values of quark and gluon fields -- the QCD condensates.
Thereby the hadron is assigned an interpolating field $\eta$ which resembles the right quantum numbers
and is built from the fundamental degrees of freedom of QCD.
On the other side, the interpolating field couples to the hadron excited from the ground state and the correlation function can be related solely to the hadronic properties for $q^2 >0$. By means of analyticity of the correlator $\Pi (q)$, dispersion relations equate both approaches. This leads to the celebrated QCD sum rules~\cite{Shifman:1978bx}.
Analysing transformed dispersion relations in a suitable range of momenta
allows a determination of hadronic properties. The generalization to nuclear matter with non-vanishing temperature or chemical potentials relies on Gibbs averaged expectation values $\expvaluePsi{\ldots}$ instead of vacuum expectation values $\expvalueZero{\ldots}$~\cite{Bochkarev:1985ex}.
In what follows we focus on the nucleon, calculate the OPE and
discuss the hadronic side; the sum rule finishes this section.

\subsection{Interpolating Fields}
\begin{sloppypar}
Following an argument of Ioffe~\cite{Ioffe:1981kw} one can write down two interpolating fields representing the
nucleon with the corresponding quantum numbers $I(J^{P})=\tfrac{1}{2} (\tfrac{1}{2}^{+} )$,
$\epsilon^{abc}[u^T_a C \gamma_\mu u_b ] \gamma_5 \gamma^\mu d_c$ and $\epsilon^{abc}[u^T_a C \sigma_{\mu\nu} u_b ] \gamma_5 \sigma^{\mu\nu} d_c \,$,
%\label{eq:interpolatingFieldsNucleon}
when restricting to fields that contain no derivatives and couple to spin $\tfrac{1}{2}$
only.\footnote{%Note, the second term can be rewritten with the identity
%$ \gamma_5 \sigma^{\alpha\beta} = \tfrac{i}{2} \epsilon^{\alpha\beta\mu\nu} \sigma_{\mu\nu} $.
Concerning conventions on metric,
Dirac and Gell-Mann matrices, charge conjugation matrix $C$ etc.\ we follow~\cite{pk:Itzykson1980}.}
Extended forms of the nucleon current may include derivatives~\cite{Braun:1992jp,Stein:1994zk}, or make use of tensor interpolating fields~\cite{Furnstahl:1995nd,Leinweber:1995fn} (also used to extrapolate the vacuum nucleon mass via QCD sum rules~\cite{pk:Langwallner2005} to (unphysical) larger values obtained on the lattice, comparable to similar efforts within chiral perturbation theory~\cite{McGovern:2006fm}). The complications in nucleon sum rules can further be dealt with when taking into consideration the coupling of positive and negative parity states to the nucleon interpolating field~\cite{Kondo:2005ur}.

In this work, our structures are always written for the proton; by exchanging $u$ and $d$ the neutron is obtained (even the neutron-proton mass difference has been analyzed in this framework~\cite{Yang:1993bp}).
As interpolating fields a Fierz rearranged and thus simplified linear combination is widely used~\cite{Furnstahl:1992pi}
\begin{equation}
\eta (x) = 2 \epsilon^{abc} \left \{ t [u^T_a(x) C \gamma_5 d_b(x) ] u_c(x) + [u^T_a(x) C d_b(x) ] \gamma_5 u_c(x) \right \} \, ,
\label{eq:interpolatingFieldNucleonGeneral1}
\end{equation}
which in the above basis reads
\begin{equation}
\label{eq:interpolatingFieldNucleonGeneralInIoffeBasis}
\tilde{\eta}(x) = \dfrac{1}{2} \epsilon^{abc} \left \{ (1-t) [u^T_a(x) C \gamma_\mu u_b(x) ] \gamma_5 \gamma^\mu d_c(x)
+ (1+t) [u^T_a(x) C \sigma_{\mu\nu} u_b(x) ] \gamma_5 \sigma^{\mu\nu} d_c(x) \right \} \, .
\end{equation}
Both currents, $\eta$ and $\tilde{\eta}$, are related by Fierz transformations whereby in such a straightforward calculation the remaining difference vanishes for symmetry reasons (analog to the exclusion of Dirac structures in~\cite{Ioffe:1981kw} when constructing all possible nucleon fields).
The consequence of these two equivalent representations~\eqref{eq:interpolatingFieldNucleonGeneral1} and~\eqref{eq:interpolatingFieldNucleonGeneralInIoffeBasis} is that two different forms of the OPE arise.
On the level of four-quark condensates the identity is not obvious and leads to relations between
different four-quark structures. There appear constraints on pure flavor
four-quark condensates which can be understood also without connection to an OPE (algebraic relations on the operator level). For mixed structures such relations do not follow. This will be discussed in section 3.
\end{sloppypar}
Our subsequent equations will be given for the ansatz~(\ref{eq:interpolatingFieldNucleonGeneral1}) with  arbitrary mixing parameter $t$.
In nucleon sum rule calculations the particular choice of the field with $t=-1$, the so-called Ioffe interpolating field, is preferred for reasons of applicability of the method and numerical stability of the evaluation procedure (cf.\ also~\cite{Leinweber:1994nm} for a discussion of an optimal nucleon interpolating field; another choice of $t$ would emphasize the negative-parity state in the sum rule~\cite{Jido:1996ia}).

\subsection{Operator Product Expansion}
Using Wilson's OPE the correlation function~(\ref{eq:correlationFunction}) can be represented asymptotically as a series of Wilson coefficients multiplied by expectation values of quark and
gluon operators, the condensates.
As outlined in appendix~\ref{ap:ope}, where also notations are summarized, these coefficients can be calculated considering quark propagation
in a gluon background field which further simplifies in the Fock-Schwinger gauge.
Application of Wick's theorem to~\eqref{eq:correlationFunction} introduces the normal ordered expectation
values, which projected on color singlets, Dirac and Lorentz scalars and restricted by the demand
for time and parity reversal invariance in cold nuclear matter leads to an expansion into local condensates.

The OPE for $\Pi (x)$ and the Fourier transform in~(\ref{eq:correlationFunction}) are important steps towards the sum rule formulation. Still the correlator can be decomposed into invariant functions.
Lorentz invariance and the requested symmetry with respect to time/parity reversal allow the decomposition
\begin{equation}
\Pi (q) = \Pi _s (q^2,qv) + \Pi _q (q^2,qv)  \slash{q} + \Pi _v (q^2,qv)  \slash{v} \, ,
\label{eq:invariantDecomposition}
\end{equation}
where $v$ is the four-velocity vector of the medium.
The three invariant functions which accordingly yield three sum rule equations can be projected out
by appropriate Dirac traces
\begin{align}
\label{eq:decomp-s}
\Pi _s (q^2,qv) =& \dfrac{1}{4} \trace{\Pi (q)} \, ,\\
\Pi _q (q^2,qv) =& \dfrac{1}{4[q^2v^2-(qv)^2]} \left \{ v^2 \trace{\slash{q} \Pi (q)} - (qv) \trace{\slash{v} \Pi (q)} \right \} \, ,\\
\label{eq:decomp-v}
\Pi _v (q^2,qv) =& \dfrac{1}{4[q^2v^2-(qv)^2]} \left \{ q^2 \trace{\slash{v} \Pi (q)} - (qv) \trace{\slash{q} \Pi (q)} \right \} \,
\end{align}
and are furthermore decomposed into even $(\mathrm{e})$ and odd $(\mathrm{o})$ parts w.r.t.\ $qv$
\begin{equation}
\label{eq:evenOddDecomposition}
\Pi_i (q^2,qv) = \Pi_i^\mathrm{e} (q^2, (qv)^2) + (qv) \Pi_i^\mathrm{o} (q^2, (qv)^2) \, .
\end{equation}
For the nucleon interpolating field~\eqref{eq:interpolatingFieldNucleonGeneral1}, this leads to
\begin{align}
\label{eq:ope-se}
\Pi_s^\mathrm{e}(q^2, (qv)^2) = &  + \dfrac{ c_1 }{16 \pi^{2} } q^2 \ln (-q^2) \condensate{ \bar{q} q } + \dfrac{3c_2}{16 \pi^2} \ln (-q^2) \condensate{ \bar{q} g_s (\sigma G) q }
\nonumber \\ & + \dfrac{2c_3}{3 \pi^2 v^2} \dfrac{(qv)^2}{q^2}  \left ( \condensate{ \bar{q} (viD)^2 q / v^2 } + \dfrac{1}{8} \condensate{ \bar{q} g_s (\sigma G) q } \right ) \, , \\
\Pi_s^\mathrm{o}(q^2, (qv)^2) = & 
- \dfrac{1}{3 v^2} \dfrac{1}{q^2} \left \{ c_1 \condensate{\bar{q} q} \condensate{\bar{q} \slash{v} q} \right \}_{\rm eff}^{1} \, , \\
\Pi_q^\mathrm{e}(q^2, (qv)^2) = & - \dfrac{c_4}{512 \pi^4} q^4 \ln (-q^2) - \dfrac{c_4}{256 \pi^2} \ln (-q^2) \condensate{ \dfrac{\alpha_s}{\pi}  G^2 } \nonumber \\
& + \dfrac{c_4}{72 \pi^2 v^2} \left ( 5 \ln (-q^2) - \dfrac{8 (qv)^2}{q^2 v^2} \right ) \condensate{ \bar{q} \slash{v} (viD) q } \nonumber
\\ & - \dfrac{c_4}{1152 \pi^2 v^2} \left ( \ln (-q^2) - \dfrac{4(qv)^2}{q^2 v^2}  \right ) \condensateAlphaPiVGsquareplusVGtsquare \nonumber \\
& - \dfrac{1}{6} \dfrac{1}{q^2} \left \{ c_1 \condensate{\bar{q} q}^2 + \dfrac{c_4}{v^2} \condensate{\bar{q} \slash{v} q}^2 \right \}_{\rm eff}^{q} \, , \\
\Pi_q^\mathrm{o}(q^2, (qv)^2) = & + \dfrac{c_4}{24 \pi^2 v^2} \ln (-q^2) \condensate{ \bar{q} \slash{v} q }
+ \dfrac{c_5}{72 \pi^2 v^2} \dfrac{1}{q^2} \condensate{ \bar{q} g_s \slash{v} (\sigma G) q } \nonumber \\
& - \dfrac{c_4}{12 \pi^2 v^2} \dfrac{1}{q^2} \left ( 1 + \dfrac{2 (qv)^2}{q^2 v^2} \right ) \left ( \condensate{ \bar{q} \slash{v} (viD)^2 q  /v^2} + \dfrac{1}{12} \condensate{ \bar{q} g_s \slash{v} (\sigma G) q } \right ) \, , \\
\Pi_v^\mathrm{e}(q^2, (qv)^2) = & + \dfrac{c_4}{12 \pi^2 v^2} q^2 \ln (-q^2) \condensate{ \bar{q} \slash{v} q }
- \dfrac{c_5}{48 \pi^2 v^2} \ln (-q^2) \condensate{ \bar{q} g_s \slash{v} (\sigma G) q } \nonumber
\\ & + \dfrac{c_4 }{2 \pi^2 v^4} \dfrac{(qv)^2}{q^2} \left ( \condensate{ \bar{q} \slash{v} (viD)^2 q  /v^2} + \dfrac{1}{12} \condensate{ \bar{q} g_s \slash{v} (\sigma G) q } \right ) \, , \\
\label{eq:ope-vo}
\Pi_v^\mathrm{o}(q^2, (qv)^2) = & + \dfrac{c_4}{288 \pi^2 v^4} \ln (-q^2) \condensateAlphaPiVGsquareplusVGtsquare
- \dfrac{5 c_4}{18 \pi^2 v^4} \ln (-q^2) \condensate{ \bar{q} \slash{v} (viD) q } \nonumber \\
& - \dfrac{1}{3 v^2} \dfrac{1}{q^2} \left \{ \dfrac{c_4}{v^2} \condensate{\bar{q} \slash{v} q}^2 \right \}_{\rm eff}^{v} \, ,
\end{align}
where the $c_i (t)$'s, being polynomials of the mixing parameter $t$, are written out below in section~\ref{sec:sumruleequations} in the final sum rules. Numerical values for condensates are collected in section~\ref{sec:numericalanalysis} where sum rules are numerically analyzed.
The contributions from four-quark condensates are written here as the usual factorized result denoted by $\{ \ldots \}_{\rm eff}^{1,q,v}$; full expressions which replace and overcome this simplification are the focus of section~\ref{sec:fourquarkcondensates} (see especially Eqs.~\eqref{eq:fqcList_s}-\eqref{eq:fqcList_v} below). Note that, in contrast to the OPE for $\Pi$ for light vector mesons with conserved currents $j_\mu=\tfrac{1}{2} (\bar{u} \gamma_\mu u \pm \bar{d} \gamma_\mu d)$ (for $\omega, \rho$ mesons), four-quark condensates enter
already without a factor $\alpha_s$ (the strong coupling) and the chiral condensate $\langle \bar{q} q \rangle$ does not appear in a renormalization invariant combination with the quark mass.

\subsection{Dispersion Relations}
The representation of the correlation function through Wilson coefficients
and condensates valid for large Euclidean momenta $q^2 < 0$ is by the provided
analiticity of $\Pi (q)$ related to the spectral density integrated over real values of the energy $q_0$.
This is reflected in the fixed-$\vec{q}$ dispersion relation of the form (up to subtractions not displayed here and written now in nuclear matter rest frame only with $q_0$ as argument)
\begin{equation}
\label{eq:dispersionRelationAnsatz}
\dfrac{1}{\pi} \int_{-\infty}^{+\infty} d\omega \dfrac{\Delta \Pi (\omega) }{\omega - q_0} = \Pi (q_0) \, ,
\end{equation}
where $\vec{q}$ is held fixed and the spectral density $\rho (\omega) = \tfrac{1}{\pi} \Delta \Pi (\omega)$ enters as the discontinuity of the correlator $\Pi$ on the real axis
\begin{equation}
\Delta \Pi (\omega) = \dfrac{1}{2 i} \lim_{\epsilon \rightarrow 0} \left [ \Pi (\omega +i\epsilon) - \Pi (\omega -i\epsilon) \right ] \, .
\end{equation}
Although dispersion relations could require polynomial subtractions enforcing convergence, such finite polynomials
vanish under Borel transformation $\mathcal{B}: f(q_0^2) \rightarrow \tilde{f}(\borel^2)$ and need not be considered here.
The correlation function is decomposed into even and odd parts using~\eqref{eq:invariantDecomposition} for $v =(1, 0,0,0)$ defining the rest frame of nuclear matter, with
\begin{align}
\Pi^{\mathrm e} (q_0^2) & \equiv \dfrac{1}{2} \left ( \Pi (q_0) + \Pi (-q_0) \right ) \mspace{20mu}= \dfrac{1}{\pi} \int_{-\infty}^{+\infty} d\omega \dfrac{{\omega \Delta} \Pi (\omega) }{\omega^2 - q_0^2} \, , \\
\Pi^{\mathrm o} (q_0^2) & \equiv \dfrac{1}{2q_0} \left ( \Pi (q_0) - \Pi (-q_0) \right ) \mspace{5mu}= \dfrac{1}{\pi} \int_{-\infty}^{+\infty} d\omega \dfrac{\Delta \Pi (\omega) }{\omega^2 - q_0^2} \, ,
\end{align}
which, given the integral representation~\eqref{eq:dispersionRelationAnsatz}, become functions of $q_0^2$.
The starting point of our sum rule analysis is the special combination
\begin{equation}
\label{eq:sumruleansatz}
\dfrac{1}{\pi} \int_{-\infty}^{+\infty} d\omega (\omega - \bar{E}) \dfrac{{ \Delta } \Pi (\omega) }{\omega^2 - q_0^2} = \Pi^{\mathrm e} (q_0^2) - \bar{E} \Pi^{\mathrm o} (q_0^2) \, ,
\end{equation}
where the l.h.s.\ encodes hadronic properties, while the r.h.s.\ is subject to the OPE~\eqref{eq:ope-se}-\eqref{eq:ope-vo}. Such a combination has been proposed in~\cite{Furnstahl:1992pi} with the motivation to suppress anti-nucleon contributions effectively by a suitable choice of the quantity $\bar{E}$.
Having in mind the usual decomposition ''resonance + continuum'', we split the l.h.s.\ of the integral into
an anti-nucleon continuum, $-\infty < \omega < \omega_-$, anti-nucleon, $\omega_- < \omega < 0$, nucleon, $0 < \omega < \omega_+$, and nucleon continuum, $\omega_+ < \omega < \infty$, and choose
\begin{equation}
\label{eq:ebardefinition}
\bar{E} = \dfrac{\int_{\omega_-}^0 d\omega \Delta \Pi (\omega) \omega e^{-\omega^2 / \borel^2}}{\int_{\omega_-}^0 d\omega \Delta \Pi (\omega) e^{-\omega^2 / \borel^2}}
\quad \text{and} \quad
 E = \dfrac{\int^{\omega_+}_0 d\omega \Delta \Pi (\omega) \omega e^{-\omega^2 / \borel^2}}{\int^{\omega_+}_0 d\omega \Delta \Pi (\omega) e^{-\omega^2 / \borel^2}} \, ,
\end{equation}
where the latter similarly defines the first moment of the Borel weighted spectral density for the positive energy
excitation.\footnote{In general, the Dirac structure of $\Delta \Pi$ would require definitions $E_i$, $\bar{E}_i$
for the distinct invariant functions $(i=s,q,v)$ of the decomposition~\eqref{eq:invariantDecomposition}. In the case considered here we assume that these weighted moments coincide with $E_{s,q,v}=E$ (analogously $\bar{E}$).
Also $\omega_{\pm}$ are simplified to be common for $s,q,v$ parts.
In the shown Borel transformed equations, decomposed terms are symbolically rearranged to full Dirac structures.}
This delivers the Borel transformed sum rule
\begin{equation}\begin{split}
\label{eq:exactFullSumRule}
(E - \bar{E}) \dfrac{1}{\pi} \int_{0}^{\omega_+} d \omega \Delta \Pi (\omega) e^{-\omega^2 / \borel^2} = \tilde{\Pi}^\mathrm{e} (\borel^2) - \dfrac{1}{\pi} \int_{\omega_+}^{\infty} d \omega
\omega \Pi^\mathrm{e}_\mathrm{per} (\omega) e^{-\omega^2 / \borel^2} \\
- \bar{E} \left \{ \tilde{\Pi}^\mathrm{o} (\borel^2) - \dfrac{1}{\pi} \int_{\omega_+}^{\infty} d \omega
\Pi^\mathrm{o}_\mathrm{per} (\omega) e^{-\omega^2 / \borel^2} \right \} 
+ \dfrac{1}{\pi} \int_{\omega_-}^{-\omega_+} d \omega \Delta \Pi (\omega) [\omega - \bar{E}] e^{-\omega^2 / \borel^2} \, .
\end{split}\end{equation}
The continuum contributions
$\Pi^\mathrm{e,o}_\mathrm{per} (\omega) \equiv \Delta \Pi (\omega) \mp \Delta \Pi (-\omega)$
are arranged on the r.h.s.\ with the reasoning of employing the semi-local duality hypothesis: For the integrated continuum the same asymptotic behavior is assumed for the correlation functions of hadronic and quark/gluon degrees of freedom in the limit of large energies. These integrals are then extended to the respective continuum thresholds $\omega_\pm$. Typically only the logarithmic terms in $\Pi$ provide discontinuities which enter the continuum integrals. To summarize, Eq.~\eqref{eq:exactFullSumRule} exhibits the typical structure of QCD sum rules: the hadronic properties on the l.h.s., i.e.\ the low-lying nucleon spectral function, are thought to be given by the operator product representation of $\Pi$ including condensates on the r.h.s.\ The last term on the r.h.s.\ accounts for asymmetric continuum thresholds, i.e.\ $\omega_- \neq -\omega_+$, and can be estimated by semi-local
quark-hadron duality.

It should be emphasized that the given sum rule is for a certain, weighted moment of a part of the nucleon spectral function.
Without further assumptions, local properties of $\Delta \Pi(\omega)$ cannot be deduced. Note also that in this form the
anti-nucleon enters inevitably the sum rule. The reasoning behind the choice of~\eqref{eq:sumruleansatz} with~\eqref{eq:ebardefinition} is that
in mean field approximation, where self-energy contributions in the propagator are real and energy-momentum
independent (cf.\ also~\cite{Furnstahl:1992pi}), the pole contribution of the nucleon propagator $G(q) = ( \slash{q} - M_N - \Sigma )^{-1}$ can be written as
\begin{equation}
G(q) = \dfrac{1}{1-\Sigma_q} \dfrac{\slash{q} + M_N^* - \slash{v} \Sigma_{v}}{(q_0 - E_+)(q_0 - E_-)}\, .
\label{eq:nucleonPropagator}
\end{equation}
Pauli corrections to positive-energy baryons and propagation of holes in the Fermi sea give rise to an additional piece $G_\mathrm{D} (q) \sim \Theta (|\vec{q_F}| - |\vec{q}\,|)$~\cite{Serot:1984ey}
vanishing for nucleon momenta $\vec{q}$ above the Fermi surface $|\vec{q_F}|$ considered here. The self-energy $\Sigma$ is decomposed into invariant structures $\Sigma = \tilde{\Sigma}_s + \Sigma_q \slash{q} + \tilde{\Sigma}_v \slash{v}$~\cite{Rusnak:1995ex} (for mean field $\Sigma_q = 0$), where one introduces scalar $\Sigma_{s} = M_N^* - M_N$ and vector self-energies $\Sigma_{v}$, which are related to the decomposition above via
$M_N^* = \tfrac{M_N + \tilde{\Sigma}_s}{1-\Sigma_q}$ and $\Sigma_v = \tfrac{\tilde{\Sigma}_v}{1-\Sigma_q}$~\cite{Furnstahl:1992pi}.
In the rest frame of nuclear matter the energy of the nucleon is $E_+$, correspondingly
$E_-$ that of the antinucleon excitation, where $E_\pm  = \Sigma_v \pm \sqrt{\vec{q}\;^{2} + M_N^{*2}}$.
Since the sum rule explicitly depends on the nucleon momentum however the self-energy $\Sigma$ as well as invariant structures $\Sigma_i$ and derived quantities acquire now a momentum dependence and become functions of the Lorentz invariants $q^2$, $qv$ and $v^2$ extending mean field theory towards the relativistic Hartree-Fock approximation~\cite{Serot:1984ey}.
Eq.~\eqref{eq:nucleonPropagator} is giving rise to a discontinuity
$\Delta G (q_0) = \tfrac{1}{2 i} \lim_{\epsilon \rightarrow 0} ( G(q_0+i\epsilon) - G(q_0-i\epsilon) )$ with a simple pole structure
\begin{equation}
\Delta G (q_0) = \dfrac{\pi}{1-\Sigma_q} \dfrac{\slash{q} + M_N^* - \slash{v} \Sigma_v}{E_+ - E_-} \left ( \delta (q_0 - E_- ) - \delta (q_0 - E_+ ) \right ) \, ,
\end{equation}
where the general expression, Eq.~\eqref{eq:ebardefinition}, identifies $\bar{E}$ with the anti-nucleon pole energy $E_-$ for all 3 Dirac structures (analogously, $E$ is identified with $E_+$). Then the l.h.s.\ of
the sum rule~\eqref{eq:exactFullSumRule} reads
\begin{equation}
\label{eq:transitionProptoGeneral}
( E_+ - E_- ) \dfrac{1}{\pi} \int_{0}^{\omega_+} d \omega \Delta \Pi (\omega) e^{-\omega^2 / \borel^2} =
- \dfrac{\lambda_N^2}{1-\Sigma_q} ( \slash{q} + M_N^* - \slash{v} \Sigma_v) e^{-E_+^2 / \borel^2}
\, .
\end{equation}
Here, $\lambda_N$ enters through the transition from Eq.~\eqref{eq:correlationFunction} to the nuclon propagator via insertion of a complete basis, retaining only the nucleon state $|q,s \rangle$ with the relation to the nucleon bispinor $\lambda_N u(p,s) = \langle \Psi |\eta (0)| q,s \rangle$ and can be combined into an effective coupling $\lambda_N^{*2}=\tfrac{\lambda_N^2}{(1-\Sigma_q)}$.
More general one can interpret Eq.~\eqref{eq:transitionProptoGeneral} as parametrization of the l.h.s.\ of~\eqref{eq:exactFullSumRule}, where integrated information of $\Delta \Pi$ is mapped onto the quantities $M_N^*$, $\Sigma_{v}$, $\lambda_N^{*2}$ and $E_\pm$, which are subject of our further analysis, by virtue of Eqs.~\eqref{eq:decomp-s}-\eqref{eq:decomp-v}
\begin{align}
\label{eq:decomposedSumRule1s}
- \lambda_N^{*2} M_N^* e^{-E_+^2 / \borel^2} & = ( E_+ - E_- ) \dfrac{1}{\pi} \int_{0}^{\omega_+} d \omega \Delta \Pi_s (\omega) e^{-\omega^2 / \borel^2} \, , \\
- \lambda_N^{*2} e^{-E_+^2 / \borel^2} & = ( E_+ - E_- ) \dfrac{1}{\pi} \int_{0}^{\omega_+} d \omega \Delta \Pi_q (\omega) e^{-\omega^2 / \borel^2} \, , \\
\label{eq:decomposedSumRule1v}
\lambda_N^{*2} \Sigma_v e^{-E_+^2 / \borel^2} & = ( E_+ - E_- ) \dfrac{1}{\pi} \int_{0}^{\omega_+} d \omega \Delta \Pi_v (\omega) e^{-\omega^2 / \borel^2} \, .
\end{align}
Due to the supposed pole structure in~\eqref{eq:nucleonPropagator} the self-energy components are related to $E_\pm$ (or more general to $E$ and $\bar{E}$) and the relations from distinct Dirac structures are coupled equations. The given general spectral integrals however not yet relate the unknown quantities, so that our numerical results presented here are not completely independent of the given nucleon propagator ansatz. These relations highlight also the dependence on the Borel mass $\borel$ which one gets rid of by averaging in an appropriate Borel window.

In~\cite{Griegel:1994xb}, it has been pointed out, that $\Pi$ also contains chiral logarithms, e.g.\ $ \overset{\mspace{12mu} \circ \mspace{3mu} 2}{m}_{\mspace{-5mu} \pi} \log \overset{\mspace{12mu} \circ \mspace{3mu} 2}{m}_{\mspace{-5mu} \pi}$, which, however, do not appear in the chiral perturbation theory expression for $M_N$. It was argued~\cite{Lee:1994hs,Birse:1996fw} that low-lying continuum like $\pi N$ excitations around $M_N$ cancel such unwanted pieces. In this respect, the parameters $M_N^*$, $\Sigma_q$, $\Sigma_v$ in~\eqref{eq:decomposedSumRule1s}-\eqref{eq:decomposedSumRule1v} are hardly to be identified with pure nucleon pole characteristics, but should be considered as measure of integrated strength of nucleon like excitations in a given interval. Moreover, many hadronic models point to a quite distributed strength or even multi-peak structures~(e.g.\ \cite{Post:2003hu}). The importance of an explicit inclusion of scattering contributions in the interval $0 \ldots \omega_+$ has been demonstrated in~\cite{Koike:1993sq,Mallik:2001ft,Adami:1992at} for finite temperature effects on the in-medium nucleon or in~\cite{pk:Morath2001} for the in-medium $D^+$ when trying to isolate the pure pole contribution.
In vacuum QCD sum rules for baryons, e.\ g.\ the nucleon, improvement of the continuum treatment is achieved by the inclusion of negative-parity states, which are equally described by a given correlation function as the corresponding positive-parity states~\cite{Jido:1996ia,Jido:1996zw,Oka:1996zz,Kondo:2005ur,Kondo:2006xz}.
Resorting to integrated strength distributions avoids these problems, but loses the tight relation to simple pole parameters.

\subsection{Sum Rule Equations}
\label{sec:sumruleequations}

Eq.~(\ref{eq:exactFullSumRule}) is the sum rule we are going to evaluate with respect to the above motivated identifications.
Inserting the decomposition~(\ref{eq:invariantDecomposition}) with~\eqref{eq:decomp-s}-\eqref{eq:ope-vo} we arrive at the three coupled sum rule equations
\begin{align}
\label{eq:sumRuleEquation_s}
\lambda_N^{*2} M_N^* e^{-(E_+^2 - \vec{q}\;^{2}) / \borel^2} &= \mspace{57mu}A_1 \borel^4 + A_2 \borel^2 + A_3 \, ,\\
\label{eq:sumRuleEquation_q}
\lambda_N^{*2} e^{-(E_+^2 - \vec{q}\;^{2}) / \borel^2} &= B_0 \borel^6 \mspace{60mu }+ B_2 \borel^2 + B_3 + B_4 / \borel^2 \, ,\\
\label{eq:sumRuleEquation_v}
\lambda_N^{*2} \Sigma_v e^{-(E_+^2 - \vec{q}\;^{2}) / \borel^2} &= \mspace{60mu }C_1 \borel^4 + C_2 \borel^2 + C_3 \, ,
\end{align}
with coefficients
\begin{align}
\label{eq:sumRuleCoefficients}
A_1 &= - \dfrac{c_1}{16\pi^2} E_1  \condensate{ \bar{q} q }  \, , \nonumber\\
A_2 &= - \dfrac{3c_2}{16\pi^2} E_0 \condensate{g_s \bar{q} \sigma G q} \, , \nonumber\\
A_3 &= -\dfrac{2c_3}{3\pi^2} \vec{q}\;^{2} \left ( \condensate{\bar{q} iD_0 iD_0 q} + \dfrac{1}{8} \condensate{g_s \bar{q} \sigma G q} \right )
-\dfrac{1}{3} E_- \left \{ c_1 \condensate{\bar{q} q} \condensate{\bar{q} \slash{v} q} \right \}_{\rm eff}^{1} \, , \nonumber\\
B_0 &= \dfrac{c_4}{256 \pi^4} E_2 \, , \nonumber\\
B_2 &= \dfrac{c_4 E_0}{24 \pi^2} E_- \condensate{q^{\dagger} q } - \dfrac{5 c_4 E_0}{72\pi^2} \condensate{ q^\dagger i D_0 q }
+ \dfrac{c_4 E_0}{256 \pi^2} \condensate{ \dfrac{\alpha_s}{\pi} G^2 } \nonumber\\
& + \dfrac{c_4 E_0}{1152 \pi^2} \condensate{ \dfrac{\alpha_s}{\pi} [(vG)^2 + (v\tilde{G})^2] } \, , \nonumber\\
B_3 &= \dfrac{c_4 \vec{q}\;^{2}}{9\pi^2} \condensate{ q^\dagger i D_0 q }
- \dfrac{c_4 \vec{q}\;^{2}}{288 \pi^2} \condensate{ \dfrac{\alpha_s}{\pi} [(vG)^2 + (v\tilde{G})^2] }
+ \dfrac{c_5 E_-}{72 \pi^2} \condensate{g_s q^\dagger \sigma G q} \nonumber\\
& - \dfrac{c_4}{4} E_- \left ( \condensate{q^\dagger iD_0 iD_0 q} + \dfrac{1}{12} \condensate{g_s q^\dagger \sigma G q} \right )
+ \dfrac{1}{6} \left \{ c_1 \condensate{\bar{q} q}^2 + \dfrac{c_4}{v^2} \condensate{\bar{q} \slash{v} q}^2 \right \}_{\rm eff}^{q} \, , \nonumber\\
B_4 &= \dfrac{c_4}{6\pi^2} \vec{q}\;^{2} \left ( \condensate{q^\dagger iD_0 iD_0 q} + \dfrac{1}{12} \condensate{g_s q^\dagger \sigma G q} \right ) \, , \nonumber\\
C_1 &= \dfrac{c_4}{12\pi^2} E_1 \condensate{ q^{\dagger} q } \, , \nonumber\\
C_2 &= \dfrac{5 c_4}{18\pi^2} E_0 E_- \condensate{ q^{\dagger} iD_0 q } 
- \dfrac{c_4 E_0}{288 \pi^2} E_- \condensate{ \dfrac{\alpha_s}{\pi} [(vG)^2 + (v\tilde{G})^2] }
- \dfrac{c_5 E_0}{48 \pi^2} \condensate{ g_s q^\dagger \sigma G q } \, , \nonumber\\
C_3 &= \dfrac{c_4}{2\pi^2} \vec{q}\;^{2} \left ( \condensate{q^\dagger iD_0 iD_0 q} + \dfrac{1}{12} \condensate{g_s q^\dagger \sigma G q} \right ) + \dfrac{1}{3} E_- \left \{ \dfrac{c_4}{v^2} \condensate{\bar{q} \slash{v} q}^2 \right \}_{\rm eff}^{v} \, ,
\end{align}
and factors $E_j$ emerging from continuum contributions, with the definition $s_0 = \omega_+^2 - \vec{q}\,^{2}$,
\begin{equation}
E_0 = \left [ 1 - e^{-s_0 / M^2} \right ], \;
E_1 = \left [ 1 - \left ( 1 + \dfrac{s_0}{M^2} \right ) e^{-s_0 / M^2} \right ], \;
E_2 = \left [ 1 - \left ( 1 + \dfrac{s_0}{M^2} + \dfrac{s_0^2}{2 M^4} \right ) e^{-s_0 / M^2} \right ] ,
\end{equation}
and the asymmetric continuum threshold integral in Eq.~\eqref{eq:exactFullSumRule} neglected.
The list \eqref{eq:sumRuleCoefficients} is exhaustive for all condensates up to mass dimension 5 in the limit of
vanishing quark masses.
The coefficients $c_i$ denote general structures due to the mixing of interpolating fields according to~\eqref{eq:interpolatingFieldNucleonGeneral1} obeying
\begin{align}
c_1 & = 7t^2 - 2t - 5, \\
c_2 & = - t^2 + 1, \\
c_3 & = 2t^2 - t - 1, \\
c_4 & = 5t^2 + 2t + 5, \\
c_5 & = 7t^2 + 10t + 7.
\end{align}

\section{Four-Quark Condensates}
\label{sec:fourquarkcondensates}
%\subsubsection*{What are Four-Quark Condensates?}
Formally, four-quark condensates are QCD ground state expectation values of hermitian products of four quark operators which are to be Dirac and Lorentz
scalars, color singlets and are to be invariant under time and parity reversal.
Thereby we restrict ourselves to equilibrated cold nuclear matter\footnote{The catalog can be extended to non-equilibrated systems lifting the demand for time reversal symmetry or to systems at finite temperature and vanishing chemical potential where charge conjugation provides a good symmetry.} but do not impose isospin symmetry from the very beginning in view of further applications, such as the proton-neutron mass difference in asymmetric cold nuclear matter (e.g.~\cite{Drukarev:2004fn}). With the following discussion of independent four-quark condensates for arbitrary numbers of flavors we allow for the inclusion of strange quark contributions as well.
Physically, the four-quark condensates quantify the correlated production of two quark-antiquark pairs in the physical vacuum.
In contrast to the square of the two-quark condensate, which accounts for uncorrelated production of two of these pairs,
the four-quark condensates are a measure of the correlation and thus evidence the complexity of the QCD ground state.
Especially, deviations from factorization, the approximation of unknown four-quark condensates in
terms of the squared chiral condensate justified in the large $N_c$ limit (cf.\ also~\cite{Leupold:2005eq}), represent effects of these more involved correlations. In this section
the classification of four-quark condensates, in the light quark sector, is performed in some detail.

\subsection{Projection and Classification}
\label{sec:fqcClassification}
The projections onto Dirac, Lorentz and color structures lead to all possible in-medium four-quark condensates
just as for the example of the non-local two-quark expectation value in appendix~\ref{ap:ope}. However the situation is even simpler since
we are only interested in the mass dimension 6 condensates, so derivatives are not required and all operators in four-quark expectation
values are to be taken at $x=0$.
%We present a classification of all possible four-quark condensates
%in medium in the case of two flavors under parity and time reversal invariance.

Using the Clifford bases $O_k \in \{ \mathbbm{1}, \gamma_\mu , \sigma_{\mu < \nu}, i\gamma_5 \gamma_\mu , \gamma_5  \}$ and
$O^m \in \{ \mathbbm{1}, \gamma^\mu , \sigma^{\mu < \nu}, i\gamma_5 \gamma^\mu , \gamma_5  \}$ which fulfill
$\trace{O_k O^m} = 4 \delta^m_k$
one can project out the Dirac indices of products of four arbitrary quark operators
\begin{equation}
\left ( \underset{\indexrm{e}}{\bar{q}_1}^{a'} \underset{\indexrm{f}}{q_2}^{a} \underset{\indexrm{g}}{\bar{q}_3}^{b'} \underset{\indexrm{h}}{q_4}^{b} \right )
= \dfrac{1}{16} \sum_{k,l=1}^{16} \left ( \bar{q}_1^{a'} O_k q_2^{a} \bar{q}_3^{b'} O^l q_4^{b} \right ) \, \underset{\indexrm{f},\indexrm{e}}{O}^k \, \underset{\indexrm{h},\indexrm{g}}{O_l} \, .
\label{eq:fqcDiracProjection}
\end{equation}
Note, here and elsewhere, Dirac indices, if explicitly shown, are attached below the concerned objects.
From~(\ref{eq:fqcDiracProjection}) there are 25 combinatorial Lorentz structures which have to be projected on condensates to obey Lorentz invariance (using the four-velocity $v_\mu$), time/parity reversal and hermiticity.
For each of the remaining 5 (10) Lorentz scalars in vacuum (medium) two possible color singlet combinations can be formed using contractions with the unity element and the generators of $SU(N_c=3)$. Thus one obtains the projection formula
\begin{equation}
\bar{q}^{a'}_1 q^a_2 \bar{q}^{b'}_3 q^b_4 = \dfrac{1}{9} \left ( \bar{q}_1 q_2 \bar{q}_3 q_4 \right ) \mathbbm{1}_{aa'} \mathbbm{1}_{bb'} + \dfrac{1}{12} \left ( \bar{q}_1 \lambda^A q_2 \bar{q}_3 \lambda^A q_4 \right ) \lambda^B_{aa'} \lambda^B_{bb'} \, .
\end{equation}
Especially, in the calculation of an operator product expansion for baryons the color condensate structures naturally arise from the product
$
\epsilon_{abc} \epsilon_{a'b'c'} \; \delta^{cc'} = \epsilon_{abc} \epsilon_{a'b'c} = \delta_{aa'} \delta_{bb'} - \delta_{ab'} \delta_{a'b}
$,
hence there the four-quark condensates generally appear in linear combinations of color structures in the form
\begin{align}
\epsilon_{abc} \epsilon_{a'b'c} \; \bar{q}^{a'}_1 q^a_2 \bar{q}^{b'}_3 q^b_4 & =  \dfrac{2}{3} \left \{ \left ( \bar{q}_1 q_2 \bar{q}_3 q_4 \right ) - \dfrac{3}{4} \left ( \bar{q}_1 \lambda^A q_2 \bar{q}_3 \lambda^A q_4 \right ) \right \} \, .
\label{eq:fqcColourDecomposition}
\end{align}

This would imply two condensate structures for each Lorentz scalar term; however, for expectation values with just one flavor (pure flavor four-quark condensates) these structures are not independent. Combining Fierz rearrangement of the Dirac contractions of pure four-quark operators with the rearrangement of the color structures, one derives the transformation equation
\begin{equation}
\left ( \bar{u} O_k \lambda^A u \bar{u} O^l \lambda^A u \right ) = - \dfrac{2}{3} \left ( \bar{u} O_k u \bar{u} O^l u \right ) - \dfrac{1}{8} \trace{O_k O_n O^l O^m} \left ( \bar{u} O_m u \bar{u} O^n u \right ) \, ,
\end{equation}
which relates the two different color combinations.
This transformation can be brought in matrix form $\vec{y} = \hat{A} \vec{x}$ with
\begin{equation}
\vec{y} = 
\begin{pmatrix}
\condensate{ \bar{q} \lambda^A q \bar{q} \lambda^A q } \\
\condensate{ \bar{q} \gamma_\alpha \lambda^A q \bar{q} \gamma^\alpha \lambda^A q } \\
\condensate{ \bar{q} \slash{v} \lambda^A q \bar{q} \slash{v} \lambda^A q } /v^2 \\
\condensate{ \bar{q} \sigma_{\alpha \beta} \lambda^A q \bar{q} \sigma^{\alpha \beta}  \lambda^A q } \\
\condensate{ \bar{q} \sigma_{\alpha \beta} \lambda^A q \bar{q} \sigma^{\gamma \delta}  \lambda^A q } g^{\alpha}_{\gamma} v^\beta v_\delta /v^2 \\
\condensate{ \bar{q} \gamma_5 \gamma_\alpha \lambda^A q \bar{q} \gamma_5 \gamma^\alpha \lambda^A q } \\
\condensate{ \bar{q} \gamma_5 \slash{v} \lambda^A q \bar{q} \gamma_5 \slash{v} \lambda^A q } /v^2 \\
\condensate{ \bar{q} \gamma_5 \lambda^A q \bar{q} \gamma_5 \lambda^A q } \\
\condensate{ \bar{q} \slash{v} \lambda^A q \bar{q} \lambda^A q } \\
\condensate{ \bar{q} \gamma_5 \gamma^\alpha \lambda^A q \bar{q} \sigma^{\beta \gamma} \lambda^A q } i \epsilon_{\alpha \beta \gamma \delta} v^\delta /2 \\
\end{pmatrix}
\mspace{20mu} \text{,} \mspace{20mu}
\vec{x} = 
\begin{pmatrix}
\condensate{ \bar{q} q \bar{q} q } \\
\condensate{ \bar{q} \gamma_\alpha q \bar{q} \gamma^\alpha q } \\
\condensate{ \bar{q} \slash{v} q \bar{q} \slash{v} q } /v^2 \\
\condensate{ \bar{q} \sigma_{\alpha \beta} q \bar{q} \sigma^{\alpha \beta}  q } \\
\condensate{ \bar{q} \sigma_{\alpha \beta} q \bar{q} \sigma^{\gamma \delta}  q }g^{\alpha}_{\gamma} v^\beta v_\delta /v^2 \\
\condensate{ \bar{q} \gamma_5 \gamma_\alpha q \bar{q} \gamma_5 \gamma^\alpha q } \\
\condensate{ \bar{q} \gamma_5 \slash{v} q \bar{q} \gamma_5 \slash{v} q } /v^2 \\
\condensate{ \bar{q} \gamma_5 q \bar{q} \gamma_5 q } \\
\condensate{ \bar{q} \slash{v}   q \bar{q}   q } \\
\condensate{ \bar{q} \gamma_5 \gamma^\alpha q \bar{q} \sigma^{\beta \gamma} q } i \epsilon_{\alpha \beta \gamma \delta} v^\delta /2 \\
\end{pmatrix} \, ,
\end{equation}
\begin{equation}
\label{eq:transformationMatrix}
\hat{A}=
\begin{pmatrix}
-7/6 & -1/2 & 0 & -1/4 & 0 & 1/2 & 0 & -1/2 & 0 & 0 \\
-2 & 1/3 & 0 & 0 & 0 & 1 & 0 & 2 & 0 & 0 \\
-1/2 & 1/2 & -5/3 & -1/4 & 1 & 1/2 & -1 & 1/2 & 0 & 0 \\
-6 & 0 & 0 & 1/3 & 0 & 0 & 0 & -6 & 0 & 0 \\
-3/2 & -1/2 & 2 & 1/4 & -2/3 & 1/2 & -2 & -3/2 & 0 & 0 \\
2 & 1 & 0 & 0 & 0 & 1/3 & 0 & -2 & 0 & 0 \\
1/2 & 1/2 & -1 & 1/4 & -1 & 1/2 & -5/3 & -1/2 & 0 & 0 \\
-1/2 & 1/2 & 0 & -1/4 & 0 & -1/2 & 0 &  -7/6 & 0 & 0 \\
0 & 0 & 0 & 0 & 0 & 0 & 0 & 0 & -5/3 & -i \\
0 & 0 & 0 & 0 & 0 & 0 & 0 & 0 & 3i & 1/3
\end{pmatrix} \, .
\end{equation}
We emphasize that the inverse transformation $\hat{A}^{-1}$ exists. However, structures for baryon sum rules typically are combinations of two color contractions, dictated by Eq.~\eqref{eq:fqcColourDecomposition}, which form components of the vector
\begin{equation}
\vec{z} = \dfrac{2}{3} \left ( \vec{x} - \dfrac{3}{4} \vec{y} \right ) = \hat{B} \vec{x} = \dfrac{2}{3} \left ( \hat{A}^{-1} - \dfrac{3}{4} \mathbbm{1} \right ) \vec{y} \, , \quad \hat{B} \equiv \tfrac{2}{3} ( \mathbbm{1} - \tfrac{3}{4} \hat{A}) \, .
\label{eq:fqcMatrixEquation}
\end{equation}
The matrix $\hat{B}$ has the fivefold eigenvalues 0 and 2, and the corresponding eigenspaces both have dimension 5, especially the kernel of $\hat{B}$ spanned by the eigenvectors to eigenvalue 0. The fact that the kernel contains more than the null vector implies that $\hat{B}$ has no inverse. The transformation of this equation into the basis of eigenvectors yields a new vector $\vec{z}\,'$ where 5 elements are to be zero. Written in components of $\vec{z}$ these relations are
\begin{align}
\label{eq:fqcConstraints1}
z_2 + z_6 & = 0 \, , \\
4 z_1 - 2 z_2 - z_4 & = 0 \, , \\
2 z_1 - z_4 + 2 z_8 & = 0 \, , \\
z_1 - z_3 - z_5 + z_7 & = 0 \, , \\
\label{eq:fqcConstraints5}
z_9 - i z_{10} & = 0 \, .
\end{align}
The first three conditions occur already in the vacuum set, the latter two ones are additional in the medium case. Of course, the conditions can be written differently, e.g., the second and third line may be
conveniently combined to $ z_1 - z_2 - z_8 = 0$ for applications.
An alternative derivation of these relations is presented in appendix~\ref{ap:constraints}.

The relations~(\ref{eq:fqcConstraints1})-(\ref{eq:fqcConstraints5}) have two important consequences: firstly, they allow to simplify pure flavor four-quark condensates in baryon sum rules;
secondly, since Eq.~(\ref{eq:fqcMatrixEquation}) can not be inverted, they forbid a direct translation from pure flavor four-quark condensates in baryon sum rules
at the order $\alpha_s^0$ to those which occur e.g.\ in sum rules for light vector mesons in the order $\alpha_s^1$.

\subsection{Four-Quark Condensates in the Nucleon QCD Sum Rule}
We have now provided all prerequisites to specify the four-quark condensates which occur in the sum rule~(\ref{eq:exactFullSumRule}). The full expressions for the four-quark condensates in the order $\alpha_s^0$, abbreviated in~\eqref{eq:sumRuleCoefficients} so far symbolically, are
\begin{align}
\label{eq:fqcList_s}
\left \{ c_1 \condensate{\bar{q} q} \condensate{\bar{q} \slash{v} q} \right \}_{\rm eff}^{1} & = \dfrac{3}{2} \epsilon_{abc} \epsilon_{a'b'c} \left ( - 2 c_2  \condensate{ \bar{u}^{a'} \slash{v} u^{a} \bar{u}^{b'} u^{b} } 
+ c_6 \condensate{ \bar{u}^{a'} \slash{v} u^{a} \bar{d}^{b'} d^{b} }
- 3 c_2 \condensate{ \bar{u}^{a'} u^{a} \bar{d}^{b'} \slash{v} d^{b} } \right .
\nonumber \\ & \left . + c_7 \condensate{ \bar{u}^{a'} \gamma_5 \gamma_\kappa u^{a} \bar{d}^{b'} \sigma_{\lambda \pi} d^{b} \epsilon^{\kappa \lambda \pi \xi} v_\xi } \right ) \, ,
\end{align}
\begin{align}
\label{eq:fqcList_q}
\left \{ c_1 \condensate{\bar{q} q}^2 + \dfrac{c_4}{v^2} \condensate{\bar{q} \slash{v} q}^2 \right \}_{\rm eff}^{q} & = \epsilon_{abc} \epsilon_{a'b'c} \left (
2 c_{9} \condensate{ \bar{u}^{a'} \gamma_\tau u^{a} \bar{u}^{b'} \gamma^\tau u^{b} } - 2 c_{9} \condensate{ \bar{u}^{a'} \slash{v} u^{a} \bar{u}^{b'} \slash{v} u^{b} / v^2 } \right .
\nonumber \\ & + 4 t \condensate{ \bar{u}^{a'} \gamma_5 \gamma_\tau u^{a} \bar{u}^{b'} \gamma_5 \gamma^\tau u^{b} } - 4 t \condensate{ \bar{u}^{a'} \gamma_5 \slash{v} u^{a} \bar{u}^{b'} \gamma_5 \slash{v} u^{b} / v^2 }
\nonumber \\ & - 9 c_2 \condensate{ \bar{u}^{a'} u^{a} \bar{d}^{b'} d^{b} }
+ \dfrac{9}{2} c_2 \condensate{ \bar{u}^{a'} \sigma_{\kappa \lambda} u^{a} \bar{d}^{b'} \sigma^{\kappa \lambda} d^{b} }
- 9 c_2 \condensate{ \bar{u}^{a'} \gamma_5 u^{a} \bar{d}^{b'} \gamma_5 d^{b} }
\nonumber \\ & + c_{10} \condensate{ \bar{u}^{a'} \gamma_\tau u^{a} \bar{d}^{b'} \gamma^\tau d^{b} } - 2 c_{9} \condensate{ \bar{u}^{a'} \slash{v} u^{a} \bar{d}^{b'} \slash{v} d^{b} / v^2 }
\nonumber \\ & \left . + c_{8} \condensate{ \bar{u}^{a'} \gamma_5 \gamma_\tau u^{a} \bar{d}^{b'} \gamma_5 \gamma^\tau d^{b} } - 4 t \condensate{ \bar{u}^{a'} \gamma_5 \slash{v} u^{a} \bar{d}^{b'} \gamma_5 \slash{v} d^{b} / v^2 } \right ) \, ,
\end{align}
\begin{align}
\label{eq:fqcList_v}
\left \{ \dfrac{c_4}{v^2} \condensate{\bar{q} \slash{v} q}^2 \right \}_{\rm eff}^{v} & = \epsilon_{abc} \epsilon_{a'b'c} \left (
 - c_{9} \condensate{ \bar{u}^{a'} \gamma_\tau u^{a} \bar{u}^{b'} \gamma^\tau u^{b} }
+ 4 c_{9} \condensate{ \bar{u}^{a'} \slash{v} u^{a} \bar{u}^{b'} \slash{v} u^{b} / v^2 } \right .
\nonumber \\ & - 2 t \condensate{ \bar{u}^{a'} \gamma_5 \gamma_\tau u^{a} \bar{u}^{b'} \gamma_5 \gamma^\tau u^{b} }
+ 8 t \condensate{ \bar{u}^{a'} \gamma_5 \slash{v} u^{a} \bar{u}^{b'} \gamma_5 \slash{v} u^{b} / v^2 }
\nonumber \\ & - c_{9} \condensate{ \bar{u}^{a'} \gamma_\tau u^{a} \bar{d}^{b'} \gamma^\tau d^{b} } + 4 c_{9} \condensate{ \bar{u}^{a'} \slash{v} u^{a} \bar{d}^{b'} \slash{v} d^{b} / v^2 }
\nonumber \\ & \left . - 2 t \condensate{ \bar{u}^{a'} \gamma_5 \gamma_\tau u^{a} \bar{d}^{b'} \gamma_5 \gamma^\tau d^{b} }
+ 8 t \condensate{ \bar{u}^{a'} \gamma_5 \slash{v} u^{a} \bar{d}^{b'} \gamma_5 \slash{v} d^{b} / v^2 } \right ) \, .
\end{align}
Here, additional polynomials which express the mixing of interpolating fields are
\begin{align}
c_6 &= t^2-2t+1,\\
c_7 &= t^2-t,\\
c_{8} &= 9t^2+10t+9,\\
c_{9} &= t^2+1,\\
c_{10} &= 11t^2+6t+11.
\end{align}
These expressions extend the non-factorized four-quark condensates for the nucleon in vacuum listed in~\cite{Koike:1993sq,Thomas:2005wc}.

\subsection{Factorization and Parametrization of Four-Quark Condensates}
Up to now we have introduced all possible four-quark condensates in the light quark sector and written out explicitly the structures which appear in
the nucleon sum rule. In such a way the sum rule equations of the type employed, e.g.\ in~\cite{Cohen:1994wm}, are equipped with complete four-quark condensates. We evaluate now the sum rule equations with the focus on these particular combinations of four-quark condensates. So we are faced with the common
problem of the poor knowledge of four-quark condensates. Usually assuming the vacuum saturation hypothesis or resorting to the large $N_c$ limit the four-quark condensates are factorized into
products of condensates with two quark operators. The factorization of four-quark condensates allows to set the proper units, however its reliability is a matter of debate. For instance, \cite{Birse:1996qp} state that the four-quark condensates in the nucleon sum rule are the expectation value of a chirally invariant operator, while $\langle \bar{q} q \rangle ^2$ is not invariant and thus a substitution by the factorized form would be inconsistent with the chiral perturbation theory expression for the nucleon self-energy.
The four-quark condensates breaking chiral symmetry might have a meaningful connection to the chiral condensate but for the chirally invariant structures such a closer relation to $\condensate{\bar{q} q}$ is not clear~\cite{Leupold:2006ih}.

Moreover, for nucleon sum rules at finite temperature $T$ (and vanishing chemical potential) it was argued in~\cite{Koike:1993sq} that the four-quark condensates are $T$ independent quite different from the behavior of $\condensate{\bar{q} q}^2$ which is why a naive factorization would lead to artificial temperature effects in the nucleon mass.

For numerical purposes it is convenient to correct the values deduced from factorization by factors $\kappa$ and
examine the effect of these correction factors on predictions from QCD sum rules. In this section the four-quark condensates classified so far in general are spelled out and the parametrization with a set of quantities $\kappa$ is defined.
In doing so one includes a density dependent factor $\kappa (n)$ in the factorized result
\begin{equation}
\condensate{ \bar{q}_{f1} \Gamma_1 \mathbbm{C}_1 q_{f1} \bar{q}_{f2} \Gamma_2 \mathbbm{C}_2 q_{f2} } = \kappa (n) \condensate{ \bar{q}_{f1} \Gamma_1 \mathbbm{C}_1 q_{f1} \bar{q}_{f2} \Gamma_2 \mathbbm{C}_2 q_{f2} }_\mathrm{fac} \, ,
\end{equation}
where $\kappa$ and the following parametrization depend on the specific condensate structure.
In linear density approximation this product ansatz obtains contributions both from the expansion
$\kappa (n) = \kappa^{(0)} + \kappa^{(1)} n$ with $\kappa^{(1)} = \tfrac{\partial \kappa (0) }{\partial n}$ and from the linearized, factorized four-quark condensate expression
$
\condensate{ \bar{q}_{f1} \Gamma_1 \mathbbm{C}_1 q_{f1} \bar{q}_{f2} \Gamma_2 \mathbbm{C}_2 q_{f2} }_\mathrm{fac} = a + b n.
$
If $\kappa^{(0)}=1$, then $\kappa^{(1)}=0$ recovers the usual factorization, which means the four-quark condensate behaves like the product of two two-quark condensates; $\kappa^{(1)}>0$ represents a stronger density dependence with respect to the factorization and vice versa. Inserting both expansions one can also describe the total density dependence of the condensates by the combination
$\medkappa{} = \kappa^{(0)} + \tfrac{a}{b} \kappa^{(1)}$,
\begin{equation}
\condensate{ \bar{q}_{f1} \Gamma_1 \mathbbm{C}_1 q_{f1} \bar{q}_{f2} \Gamma_2 \mathbbm{C}_2 q_{f2} } =
a \kappa^{(0)} + b \medkappa{} n
\end{equation}
such that for $\medkappa{}=0$ the condensate is (in first order) independent of density. For condensates with vanishing $a$ or $b$ in factorization we choose $a=\vaccondensate{ \bar{q} q}^2$ and $b=\vaccondensate{\bar{q} q} \sigma_{\rm N} / m_q$ as scale to study deviations from zero and denote these instances by $\tkappa{}$.
The classification of possible four-quark condensates is collected together with the specific $\kappa$ parametrization in Tabs.~\ref{tab:listNonMixingFQC} and~\ref{tab:listMixingFQC}.

\subsubsection*{Non-Flavor Mixing Case}
The condensates which contain only one flavor are listed in Tab.~\ref{tab:listNonMixingFQC}.
From the demand for parity and time reversal invariance only $5$~$(10)$ Dirac and Lorentz scalar four quark operators remained in vacuum (medium). Further, these structures carry color indices and must be projected on colorless objects for which there are two ways. However, since the same flavors occur, both color combinations can be alternatively rearranged
via Fierz transformation. Hence, there are only $5$~$(10)$ independent $\kappa$ parameter sets in the Tab.~\ref{tab:listNonMixingFQC}, although both color alternatives are listed. The parameter sets with indices $1,2$ are related by the transformation~\eqref{eq:transformationMatrix}.

\subsubsection*{Flavor Mixing Case}
Here the condensates containing two quark operator pairs are distinguished by flavor.
The numbering is as for the pure flavor structures. However, the conversion of the two color contractions is not possible due to 
different flavors. Compared to the non-flavor mixing case the missing exchange symmetry of $\bar{q} q$ contractions due to different flavors allows additional placements of Dirac matrices and thus leads to 4 additional condensate structures in medium (see Tab.~\ref{tab:listMixingFQC}). Therefore, 10 (24) flavor-mixed four quark condensates and thus $\kappa$ parameter pairs appear in vacuum (medium).

\paragraph{}
Hence, there exist in medium \{vacuum\} for $n_f$ flavors without flavor symmetry taken into account $2n_f (6n_f-1)$ \{$5n_f^2$\} independent four-quark condensates being Lorentz invariant expectation values of hermitian products of four quark operators constrained by time and parity reversal invariance. Symmetry under flavor rotation reduces these numbers to $20$ \{$10$\}, respectively. Finally note that these are also the numbers of necessary $\medkappa{}$ parameters. Since the four-quark condensates in operator product expansions obtained from the medium projections in the limit of vanishing baryon density $n$ should coincide with the vacuum result, this leads by contraction of vacuum and medium projections of four-quark condensates to the relations
$\vackappa{v',t',a'} = \tfrac{1}{4} \vackappa{v,t,a}$, which have already been included in Tabs.~\ref{tab:listNonMixingFQC} and~\ref{tab:listMixingFQC}. Further, Lorentz projections which exist only in medium imply no new $\vackappa{}$ parameters and so the number of $\medkappa{}$ in medium reduces consistently to the number of $\vackappa{}$ and four-quark condensates in vacuum.

\begin{table}[htb]
\begin{center}
\begin{tabular}{clr}
\hline
\\
Indices & Full condensate & Parametrized Factorization\\
& &in Linear Density Approximation\\
\hline
\\
$ {\rm 1s} $ & $\condensate{ \bar{u} u \bar{u} u }$ & $ \tfrac{11}{12} \left ( \vackappa{1s} \vaccondensate{ \bar{q} q }^2 + \medkappa{1s} n \xi \right ) $ \\

$ {\rm 1v} $ & $\condensate{ \bar{u} \gamma_\alpha u \bar{u} \gamma^\alpha u }$ & $ - \tfrac{1}{3} \left ( \vackappa{1v} \vaccondensate{ \bar{q} q }^2 + \medkappa{1v} n \xi \right ) $ \\

$ {\rm 1v'} $ & $\condensate{ \bar{u} \slash{v} u \bar{u} \slash{v} u } /v^2$ & $ - \tfrac{1}{12} \left ( \tfrac{1}{4} \vackappa{1v} \vaccondensate{ \bar{q} q }^2 + \medkappa{1v'} n \xi \right ) $ \\

$ {\rm 1t} $ & $\condensate{ \bar{u} \sigma_{\alpha \beta} u \bar{u} \sigma^{\alpha \beta}  u }$ & $ - \left ( \vackappa{1t} \vaccondensate{ \bar{q} q }^2 + \medkappa{1t} n \xi \right ) $ \\

$ {\rm 1t'} $ & $\condensate{ \bar{u} \sigma_{\alpha \beta} u \bar{u} \sigma^{\gamma \delta}  u }g^{\alpha}_{\gamma} v^\beta v_\delta /v^2$ & $ - \tfrac{1}{4} \left ( \tfrac{1}{4} \vackappa{1t} \vaccondensate{ \bar{q} q }^2 + \medkappa{1t'} n \xi \right ) $ \\

$ {\rm 1a} $ & $\condensate{ \bar{u} \gamma_5 \gamma_\alpha u \bar{u} \gamma_5 \gamma^\alpha u }$ & $ \tfrac{1}{3} \left ( \vackappa{1a} \vaccondensate{ \bar{q} q }^2 + \medkappa{1a} n \xi \right ) $ \\

$ {\rm 1a'} $ & $\condensate{ \bar{u} \gamma_5 \slash{v} u \bar{u} \gamma_5 \slash{v} u } /v^2$ & $ \tfrac{1}{12} \left ( \tfrac{1}{4} \vackappa{1a} \vaccondensate{ \bar{q} q }^2 + \medkappa{1a'} n \xi \right ) $ \\

$ {\rm 1p} $ & $\condensate{ \bar{u} \gamma_5 u \bar{u} \gamma_5 u }$ & $ - \tfrac{1}{12} \left ( \vackappa{1p} \vaccondensate{ \bar{q} q }^2 + \medkappa{1p} n \xi \right ) $ \\

$ {\rm 1vs} $ & $\condensate{ \bar{u} \slash{v} u \bar{u} u }$ & $ \tmedkappa{1vs} n \xi $ \\

$ {\rm 1at} $ & $\condensate{ \bar{u} \gamma_5 \gamma_\kappa u \bar{u} \sigma_{\lambda \pi} u } \epsilon^{\kappa \lambda \pi \xi} v_\xi $ & $ \tmedkappa{1at} n \xi $ \\

\\

$ {\rm 2s} $ & $\condensate{ \bar{u} \lambda^A u \bar{u} \lambda^A u }$ & $ - \tfrac{4}{9} \left ( \vackappa{2s} \vaccondensate{ \bar{q} q }^2 + \medkappa{2s} n \xi \right ) $ \\

$ {\rm 2v} $ & $\condensate{ \bar{u} \gamma_\alpha \lambda^A u \bar{u} \gamma^\alpha \lambda^A u }$ & $ - \tfrac{16}{9} \left ( \vackappa{2v} \vaccondensate{ \bar{q} q }^2 + \medkappa{2v} n \xi \right ) $ \\

$ {\rm 2v'} $ & $\condensate{ \bar{u} \slash{v} \lambda^A u \bar{u} \slash{v} \lambda^A u } /v^2$ & $ - \tfrac{4}{9} \left ( \tfrac{1}{4} \vackappa{2v} \vaccondensate{ \bar{q} q }^2 + \medkappa{2v'} n \xi \right ) $ \\

$ {\rm 2t} $ & $\condensate{ \bar{u} \sigma_{\alpha \beta} \lambda^A u \bar{u} \sigma^{\alpha \beta}  \lambda^A u }$ & $ - \tfrac{16}{3} \left ( \vackappa{2t} \vaccondensate{ \bar{q} q }^2 + \medkappa{2t} n \xi \right ) $ \\

$ {\rm 2t'} $ & $\condensate{ \bar{u} \sigma_{\alpha \beta} \lambda^A u \bar{u} \sigma^{\gamma \delta}  \lambda^A u }g^{\alpha}_{\gamma} v^\beta v_\delta /v^2$ & $ - \tfrac{4}{3} \left ( \tfrac{1}{4} \vackappa{2t} \vaccondensate{ \bar{q} q }^2 + \medkappa{2t'} n \xi \right ) $ \\

$ {\rm 2a} $ & $\condensate{ \bar{u} \gamma_5 \gamma_\alpha \lambda^A u \bar{u} \gamma_5 \gamma^\alpha \lambda^A u }$ & $ \tfrac{16}{9} \left ( \vackappa{2a} \vaccondensate{ \bar{q} q }^2 + \medkappa{2a} n \xi \right ) $ \\

$ {\rm 2a'} $ & $\condensate{ \bar{u} \gamma_5 \slash{v} \lambda^A u \bar{u} \gamma_5 \slash{v} \lambda^A u } /v^2$ & $ \tfrac{4}{9} \left ( \tfrac{1}{4} \vackappa{2a} \vaccondensate{ \bar{q} q }^2 + \medkappa{2a'} n \xi \right ) $ \\

$ {\rm 2p} $ & $\condensate{ \bar{u} \gamma_5 \lambda^A u \bar{u} \gamma_5 \lambda^A u }$ & $ - \tfrac{4}{9} \left ( \vackappa{2p} \vaccondensate{ \bar{q} q }^2 + \medkappa{2p} n \xi \right ) $ \\

$ {\rm 2vs} $ & $\condensate{ \bar{u} \slash{v} \lambda^A u \bar{u} \lambda^A u }$ & $ \tmedkappa{2vs} n \xi $ \\

$ {\rm 2at} $ & $\condensate{ \bar{u} \gamma_5 \gamma_\kappa \lambda^A u \bar{u} \sigma_{\lambda \pi} \lambda^A u } \epsilon^{\kappa \lambda \pi \xi} v_\xi $ & $ \tmedkappa{2at} n \xi $ \\

\\
\hline
\end{tabular}
\end{center}
\caption{Two complete sets (indices 1 and 2) of independent non-flavor-mixing four-quark condensates differing in color structure and their parametrization with $\kappa$ in strict linear density approximation $(\xi = \vaccondensate{ \bar{q} q} \sigma_{\rm N} / m_{\rm q})$. The sets are related by a Fierz transformation. A similar table for flavor $d$ instead $u$ appears for an exhaustive list of four-quark condensates for the two-flavor case $n_f =2$.}
\label{tab:listNonMixingFQC}
\end{table}

\begin{table}[htbp]
\begin{center}
\begin{tabular}{clr}
\hline
\\
Indices & Full condensate & Parametrized Factorization\\
& &in Linear Density Approximation\\
\hline
\\
$ {\rm 3s} $ & $\condensate{ \bar{u} u \bar{d} d }$ & $  \vackappa{3s} \vaccondensate{ \bar{q} q }^2 + \medkappa{3s} n \xi $ \\

$ {\rm 3v} $ & $\condensate{ \bar{u} \gamma_\alpha u \bar{d} \gamma^\alpha d }$ & $ \tvackappa{3v} \vaccondensate{ \bar{q} q }^2 + \tmedkappa{3v} n \xi  $ \\

$ {\rm 3v'} $ & $\condensate{ \bar{u} \slash{v} u \bar{d} \slash{v} d } /v^2$ & $ \tfrac{1}{4} \tvackappa{3v} \vaccondensate{ \bar{q} q }^2 + \tmedkappa{3v'} n \xi  $ \\

$ {\rm 3t} $ & $\condensate{ \bar{u} \sigma_{\alpha \beta} u \bar{d} \sigma^{\alpha \beta}  d }$ & $ \tvackappa{3t} \vaccondensate{ \bar{q} q }^2 + \tmedkappa{3t} n \xi  $ \\

$ {\rm 3t'} $ & $\condensate{ \bar{u} \sigma_{\alpha \beta} u \bar{d} \sigma^{\gamma \delta}  d }g^{\alpha}_{\gamma} v^\beta v_\delta /v^2$ & $ \tfrac{1}{4} \tvackappa{3t} \vaccondensate{ \bar{q} q }^2 + \tmedkappa{3t'} n \xi  $ \\

$ {\rm 3a} $ & $\condensate{ \bar{u} \gamma_5 \gamma_\alpha u \bar{d} \gamma_5 \gamma^\alpha d }$ & $ \tvackappa{3a} \vaccondensate{ \bar{q} q }^2 + \tmedkappa{3a} n \xi  $ \\

$ {\rm 3a'} $ & $\condensate{ \bar{u} \gamma_5 \slash{v} u \bar{d} \gamma_5 \slash{v} d } /v^2$ & $ \tfrac{1}{4} \tvackappa{3a} \vaccondensate{ \bar{q} q }^2 + \tmedkappa{3a'} n \xi  $ \\

$ {\rm 3p} $ & $\condensate{ \bar{u} \gamma_5 u \bar{d} \gamma_5 d }$ & $ \tvackappa{3p} \vaccondensate{ \bar{q} q }^2 + \tmedkappa{3p} n \xi  $ \\

$ {\rm 3vs} $ & $\condensate{ \bar{u} \slash{v} u \bar{d} d }$ & $ \medkappa{3vs}  \vaccondensate{ \bar{q} q } 3 n/2  $ \\

$ {\rm 3at} $ & $\condensate{ \bar{u} \gamma_5 \gamma_\kappa u \bar{d} \sigma_{\lambda \pi} d } \epsilon^{\kappa \lambda \pi \xi} v_\xi $ & $ \medkappa{3at}  \vaccondensate{ \bar{q} q } 3 n/2  $ \\

\\

$ {\rm 4s} $ & $\condensate{ \bar{u} \lambda^A u \bar{d} \lambda^A d }$ & $ \tvackappa{4s} \vaccondensate{ \bar{q} q }^2 + \tmedkappa{4s} n \xi $ \\

$ {\rm 4v} $ & $\condensate{ \bar{u} \gamma_\alpha \lambda^A u \bar{d} \gamma^\alpha \lambda^A d }$ & $  \tvackappa{4v} \vaccondensate{ \bar{q} q }^2 + \tmedkappa{4v} n \xi  $ \\

$ {\rm 4v'} $ & $\condensate{ \bar{u} \slash{v} \lambda^A u \bar{d} \slash{v} \lambda^A d } /v^2$ & $ \tfrac{1}{4} \tvackappa{4v} \vaccondensate{ \bar{q} q }^2 + \tmedkappa{4v'} n \xi $ \\

$ {\rm 4t} $ & $\condensate{ \bar{u} \sigma_{\alpha \beta} \lambda^A u \bar{d} \sigma^{\alpha \beta}  \lambda^A d }$ & $ \tvackappa{4t} \vaccondensate{ \bar{q} q }^2 + \tmedkappa{4t} n \xi $ \\

$ {\rm 4t'} $ & $\condensate{ \bar{u} \sigma_{\alpha \beta} \lambda^A u \bar{d} \sigma^{\gamma \delta}  \lambda^A d }g^{\alpha}_{\gamma} v^\beta v_\delta /v^2$ & $ \tfrac{1}{4} \tvackappa{4t} \vaccondensate{ \bar{q} q }^2 + \tmedkappa{4t'} n \xi $ \\

$ {\rm 4a} $ & $\condensate{ \bar{u} \gamma_5 \gamma_\alpha \lambda^A u \bar{d} \gamma_5 \gamma^\alpha \lambda^A d }$ & $ \tvackappa{4a} \vaccondensate{ \bar{q} q }^2 + \tmedkappa{4a} n \xi $ \\

$ {\rm 4a'} $ & $\condensate{ \bar{u} \gamma_5 \slash{v} \lambda^A u \bar{d} \gamma_5 \slash{v} \lambda^A d } /v^2$ & $ \tfrac{1}{4} \tvackappa{4a} \vaccondensate{ \bar{q} q }^2 + \tmedkappa{4a'} n \xi $ \\

$ {\rm 4p} $ & $\condensate{ \bar{u} \gamma_5 \lambda^A u \bar{d} \gamma_5 \lambda^A d }$ & $ \tvackappa{4p} \vaccondensate{ \bar{q} q }^2 + \tmedkappa{4p} n \xi $ \\

$ {\rm 4vs} $ & $\condensate{ \bar{u} \slash{v} \lambda^A u \bar{d} \lambda^A d }$ & $ \tmedkappa{4vs} \vaccondensate{ \bar{q} q } 3 n/2  $ \\

$ {\rm 4at} $ & $\condensate{ \bar{u} \gamma_5 \gamma_\kappa \lambda^A u \bar{d} \sigma_{\lambda \pi} \lambda^A d } \epsilon^{\kappa \lambda \pi \xi} v_\xi $ & $ \medkappa{4at}  \vaccondensate{ \bar{q} q } 3 n/2  $ \\

\\

$ {\rm 5vs} $ & $\condensate{ \bar{d} \slash{v} d \bar{u} u }$ & $ \medkappa{5vs}  \vaccondensate{ \bar{q} q } 3 n/2  $ \\

$ {\rm 5at} $ & $\condensate{ \bar{d} \gamma_5 \gamma_\kappa d \bar{u} \sigma_{\lambda \pi} u } \epsilon^{\kappa \lambda \pi \xi} v_\xi $ & $ \medkappa{5at}  \vaccondensate{ \bar{q} q } 3 n/2  $ \\

\\

$ {\rm 6vs} $ & $\condensate{ \bar{d} \slash{v} \lambda^A d \bar{u} \lambda^A u }$ & $ \tmedkappa{6vs} \vaccondensate{ \bar{q} q } 3 n/2  $ \\

$ {\rm 6at} $ & $\condensate{ \bar{d} \gamma_5 \gamma_\kappa \lambda^A d \bar{u} \sigma_{\lambda \pi} \lambda^A u } \epsilon^{\kappa \lambda \pi \xi} v_\xi $ & $ \medkappa{6at}  \vaccondensate{ \bar{q} q } 3 n/2  $ \\

\\
\hline
\end{tabular}
\end{center}
\caption{A complete set of independent flavor-mixing four-quark condensates and their parametrization by $\kappa$ parameters in strict linear density approximation. Additional parameters (indices $5$ and $6$) are required for structures which cannot be exchanged.}
\label{tab:listMixingFQC}
\end{table}
\clearpage

Insertion of these parametrization into the relevant sums of four-quark condensates~\eqref{eq:fqcList_s}-\eqref{eq:fqcList_v} yields effective $\kappa$ parameters as linear combinations of the previously defined condensate-specific parameters. The sum rule is only sensitive to these effective combinations and can thus only reveal information on the behavior of specific linear combinations of four-quark condensates. Therefore in the sum rule analysis the three parameters $\medkappa{s}$, $\medkappa{q}$, $\tmedkappa{v}$ describing the density dependence enter as
\begin{align}
\label{eq:fqcparametrization_combination_s}
\left \{ c_1 \condensate{\bar{q} q} \condensate{\bar{q} \slash{v} q} \right \}_{\rm eff}^{1} & = c_1 \left (
\medkappa{s} \vaccondensate{ \bar{q} q} \dfrac{3}{2} n \right ) \, ,\\
\label{eq:fqcparametrization_combination_q}
\left \{ c_1 \condensate{\bar{q} q}^2 + \dfrac{c_4}{v^2} \condensate{\bar{q} \slash{v} q}^2 \right \}_{\rm eff}^{q}  & = c_1 \left (
\vackappa{q} \vaccondensate{ \bar{q} q }^2 + \medkappa{q} \vaccondensate{ \bar{q} q} \dfrac{\sigma_{\rm N}}{ m_{\rm q}} n \right ) \, , \\
\label{eq:fqcparametrization_combination_v}
\left \{ \dfrac{c_4}{v^2} \condensate{\bar{q} \slash{v} q}^2 \right \}_{\rm eff}^{v} & = c_4 \left ( \tmedkappa{v} \vaccondensate{ \bar{q} q} \dfrac{\sigma_{\rm N}}{ m_{\rm q}} n \right ) \,
\end{align}
and are functions of the mixing angle $t$ as well. However, we restrict this discussion to the limit of the Ioffe interpolating field $t=-1$. Note again, the $\medkappa{}$ values are effective combinations representing the density dependence of the respective condensate lists~\eqref{eq:fqcList_s}-\eqref{eq:fqcList_v} and thus negative $\medkappa{}$, a four-quark condensate behavior contrary to the factorization assumption, comprise cancellation effects within these condensate combinations.

\subsubsection*{Density dependence of four-quark condensates from models}
It is instructive to derive values for the effective density dependence parameters $\medkappa{}$.
Expectation values of four-quark operators in the nucleon were previously calculated in a perturbative chiral quark model~\cite{Drukarev:2003xd} and taken into account in sum rule evaluations for the in-medium nucleon~\cite{Drukarev:2004zg}. (Corrections to the factorization of four-quark condensates in nucleon sum rules have also been considered in the framework of the Nambu-Jona-Lasinio model in~\cite{Celenza:1994ri}. Lattice evaluations of four-quark operators in the nucleon are yet restricted to combinations which avoid the mixing with lower dimensional operators on the lattice~\cite{Gockeler:2001xw}, and provide not yet enough information to constrain Eqs.~\eqref{eq:fqcList_s}-\eqref{eq:fqcList_v}.) The results in~\cite{Drukarev:2003xd} can be translated to our $\kappa$ parameters. However, only such color combinations being significant in baryon sum rules are considered, see left column in Tab.~\ref{tab:condensatePartsFromDrukarev}.
We note that the values given in~\cite{Drukarev:2003xd} have to be corrected slightly in order to reach full consistency with the Fierz relations~\eqref{eq:fqcConstraints1}-\eqref{eq:fqcConstraints5}, which are an operator identity and thus must be fulfilled also for expectation values in the nucleon. An optimized minimally corrected set is found by the following procedure:
minimize the relative deviation of all separate values compared to values delivered in the parametrization of~\cite{Drukarev:2003xd}
(this is in the order of 10 \%, however with different possible adjustments);
from these configurations choose the set with smallest sum of separate deviations (this deviation
sum estimates to 40 \% and different configurations are close to this value).
The results from which the relevant density dependence for our condensate classification is obtained are collected in Tab.~\ref{tab:condensatePartsFromDrukarev}; our slight modifications of values in the original parametrization~\cite{Drukarev:2003xd} are documented in Tabs.~\ref{tab:parametersDrukarevPureFlavor} and~\ref{tab:parametersDrukarevMixedFlavor} in appendix~\ref{ap:expectationValues}.

\begin{table}[htbp]
\begin{center}
\begin{tabular}{|c|c|}
\hline
Mean Nucleon Matrix Element & PCQM model $[\vaccondensate{\bar{q} q}]$ \\
(to be color contracted with $\epsilon_{abc} \epsilon_{a'b'c'}$) & \\
\hline
$   \condensate{\bar{u}^{a'} u^a \bar{u}^{b'} u^b}_N$ & $ 3.993 $ \\
$   \condensate{\bar{u}^{a'} \gamma_\alpha u^a \bar{u}^{b'} \gamma^\alpha u^b}_N$ & $ 1.977 $ \\
$   \condensate{\bar{u}^{a'} \slash{v} u^a \bar{u}^{b'} \slash{v} u^b}_N /v^2$ & $ 0.432 $ \\
$   \condensate{\bar{u}^{a'} \sigma_{\alpha \beta} u^a \bar{u}^{b'} \sigma^{\alpha \beta} u^b}_N$ & $ 12.024 $ \\
$   \condensate{\bar{u}^{a'} \sigma_{\alpha \beta} u^a \bar{u}^{b'} \sigma^{\alpha \delta} u^b}_N v^\beta v_\delta /v^2$ & $ 3.045 $ \\
$   \condensate{\bar{u}^{a'} \gamma_5 \gamma_\alpha u^a \bar{u}^{b'} \gamma_5 \gamma^\alpha u^b}_N$ & $ -1.980 $ \\
$   \condensate{\bar{u}^{a'} \gamma_5 \slash{v} u^a \bar{u}^{b'} \gamma_5 \slash{v} u^b}_N /v^2$ & $ -0.519 $ \\
$   \condensate{\bar{u}^{a'} \gamma_5 u^a \bar{u}^{b'} \gamma_5 u^b}_N$ & $ 2.016 $ \\
$   \condensate{\bar{u}^{a'} \slash{v} u^a \bar{u}^{b'} u^b}_N$ & $ - $ \\
$   \condensate{\bar{u}^{a'} \gamma_5 \gamma_\kappa u^a \bar{u}^{a'} \sigma_{\lambda \pi} u^b}_N \epsilon^{\kappa \lambda \pi \xi} v_\xi$ & $ - $ \\
& \\
$   \condensate{\bar{u}^{a'} u^a \bar{d}^{b'} d^b}_N$ & $ 3.19 $ \\
$   \condensate{\bar{u}^{a'} \gamma_\alpha u^a \bar{d}^{b'} \gamma^\alpha d^b}_N$ & $ -2.05 $ \\
$   \condensate{\bar{u}^{a'} \slash{v} u^a \bar{d}^{b'} \slash{v} d^b}_N /v^2$ & $ -0.73 $ \\
$   \condensate{\bar{u}^{a'} \sigma_{\alpha \beta} u^a \bar{d}^{b'} \sigma^{\alpha \beta} d^b}_N$ & $ 3.36 $ \\
$   \condensate{\bar{u}^{a'} \sigma_{\alpha \beta} u^a \bar{d}^{b'} \sigma^{\alpha \delta} d^b}_N v^\beta v_\delta /v^2$ & $ 1.11 $ \\
$   \condensate{\bar{u}^{a'} \gamma_5 \gamma_\alpha u^a \bar{d}^{b'} \gamma_5 \gamma^\alpha d^b}_N$ & $ 1.66 $ \\
$   \condensate{\bar{u}^{a'} \gamma_5 \slash{v} u^a \bar{d}^{b'} \gamma_5 \slash{v} d^b}_N /v^2$ & $ 0.37 $ \\
$   \condensate{\bar{u}^{a'} \gamma_5 u^a \bar{d}^{b'} \gamma_5 d^b}_N$ & $ -0.185 $ \\
$   \condensate{\bar{u}^{a'} \slash{v} u^a \bar{d}^{b'} d^b}_N$ & $ -0.245 $ \\
$   \condensate{\bar{u}^{a'} \gamma_5 \gamma_\kappa u^a \bar{d}^{b'} \sigma_{\lambda \pi} d^b}_N \epsilon^{\kappa \lambda \pi \xi} v_\xi$ & $ - $ \\
\hline
\end{tabular}
\end{center}
\caption{The combinations arranged as in the vector $\vec{z}$ of four-quark expectation values obtained from the (partially modified) set taken from a perturbative chiral quark model calculation (PCQM) in~\cite{Drukarev:2003xd} from which the characteristic density dependence of four-quark condensates, the value of
$\medkappa{}$, is derived. Isospin symmetry $N=\tfrac{1}{2}(p+n)$ of the nuclear matter ground state is assumed. The values in the pure flavor sector (upper part) are tuned to obey Fierz relations~\eqref{eq:fqcConstraints1}-\eqref{eq:fqcConstraints5} on the accuracy level $ < 0.01 \vaccondensate{\bar{q} q}$. For three combinations no results are provided in~\cite{Drukarev:2003xd} as indicated by ''$-$''.}
\label{tab:condensatePartsFromDrukarev}
\end{table}

The connection to our $\kappa$ parameters is derived as follows: Generally, in linear density approximation
condensates behave like
$
\expvalue{A} = \expvalueZero{A} + n \expvalueNucleon{A}
$.
If one compares our parametrized density dependent part of each four-quark condensate with the evaluation of nucleon matrix elements of four-quark operators in the combinations in Tab.~\ref{tab:condensatePartsFromDrukarev} one obtains values for linear combinations of $\kappa$ parameters. The linear combinations refer to the two distinct color alternatives representing, as mentioned above, the typical color combination in baryon sum rules. These values can thus be applied to give also the required effective parameters\footnote{ Note some difference to the OPE part stated in eqations (87)-(89) of~\cite{Drukarev:2004zg} for the whole combination of the density dependent four-quark condensate contribution. Our equivalent OPE calculation utilizing the same nucleon four-quark expectation values (encoded in $\medkappa{s,q}$, $\tmedkappa{v}$ as above) yields $\Pi_{\mathrm 4q} = ( 0.49 \tfrac{(qp)}{M_N} \mathbbm{1} + 0.52 \slash{q} + 0.57 \tfrac{(qp)}{M_N^2} \slash{v} ) \tfrac{\condensate{\bar{q}q}}{q^2} n $, with $p=M_N v$. } (apart from the term $\sim \condensate{\bar{u}^{a'} \gamma_5 \gamma_\kappa u^a \bar{d}^{b'} \sigma_{\lambda \pi} d^b}_N \epsilon^{\kappa \lambda \pi \xi} v_\xi$ not considered in~\cite{Drukarev:2003xd} which we had to neglect in the determination of $\medkappa{s}$)
\begin{equation}
\label{eq:pcqmKappaSet}
\medkappa{s} = -0.25 \, , \quad \medkappa{q} = -0.10 \, , \quad \tmedkappa{v} = -0.03 \, .
\end{equation}
Note that individual $\medkappa{}$ parameters are not small compared to these effective numbers indicating significant cancellation effects in the density dependent parts of combined four-quark condensates. Moreover, for pure flavor four-quark condensates the ambiguity due to Fierz relations between operators does not allow to prefer a specific condensate type as dominating the four-quark condensates in the sum rule. For further attempts to gain estimates of four-quark condensates we refer the interested reader to~\cite{Zschocke:2005gr}.

\subsubsection*{Remark on Chiral Symmetry}
The chiral condensate is often considered an order parameter for the $SU(n_f)_A$ chiral symmetry of QCD. Its change however might partially originate from virtual low-momentum pions and thus could not clearly signal partial restoration of chiral symmetry in matter~\cite{Birse:1996fw}. The interpretation of four-quark condensates as order parameters for spontaneous break-down of chiral symmetry is an open issue. In~\cite{Leupold:2006ih} a specific combination of four-quark condensates arising from the difference between vector and axial-vector correlators is proposed as such an alternative parameter. This combination (shown to agree with vacuum factorization in the analysis of tau-lepton decay data~\cite{Bordes:2005wv}) is distinct from the above four-quark condensate lists in the nucleon channel as well as from the combination in the $\omega$ sum rule~\cite{Thomas:2005dc}.
For instance, in vacuum nucleon QCD sum rules the four-quark condensate combination (the vacuum limit of Eq.~\eqref{eq:fqcList_q} with isospin symmetry being applied; $\psi$ is the flavor vector) enters as sum of a chirally invariant part
\begin{equation}
\left [ 2(2t^2+t+2) \condensate{\bar{\psi} \gamma_\mu \psi \bar{\psi} \gamma^\mu \psi}
+ (3t^2+4t+3) \condensate{\bar{\psi} \gamma_5 \gamma_\mu \psi \bar{\psi} \gamma_5 \gamma^\mu \psi} \right ] - \dfrac{3}{4} \left [\text{color structures with} \, \lambda^A \right ] \nonumber \, ,
\end{equation}
and a part which breaks this symmetry (pointed out in factorized form already in~\cite{Jido:1996ia})
\begin{equation}
\left [ 3(t^2-1) \left ( \condensate{\bar{\psi} \psi \bar{\psi} \psi} + \condensate{\bar{\psi} \gamma_5 \psi \bar{\psi} \gamma_5 \psi} - \dfrac{1}{2} \condensate{\bar{\psi} \sigma_{\mu \nu} \psi \bar{\psi} \sigma^{\mu \nu} \psi} \right ) \right ] - \dfrac{3}{4} \left [\text{color structures with} \, \lambda^A \right ] \nonumber \, .
\end{equation}
In the preferred case $t=-1$ only the chirally invariant part survives and thus this remainder cannot be used as an order parameter.
Additional insight into the change of four-quark condensates and their role as order parameters of spontaneous chiral symmetry breaking could be acquired from other hadronic channels, as the generalization of further baryon sum rules in vacuum~\cite{Leinweber:1989hh,Lee:2002jb,Lee:2006bu} to the medium case, e.g.\ for the $\Delta$~\cite{Jin:1994vw,Johnson:1995sk}.

\section{Analysis}
\subsection{Approximations}
In text book examples of QCD sum rules for light vector mesons~(e.g.\ \cite{Reinders:1984sr,pk:Narison2004}) one usually considers mass equations and optimizes them for maximum flatness w.r.t.\ the Borel mass. This, however, includes often derivative sum rules and seems not to be appropriate in the case of fermions where the condensates are distributed over coupled sum rule equations for several invariant functions. Despite of this, equations for the self-energies can be formed dividing Eqs.~\eqref{eq:sumRuleEquation_s} and~\eqref{eq:sumRuleEquation_v} by~\eqref{eq:sumRuleEquation_q} thus arriving at a generalization of Ioffe's formula~\cite{Ioffe:1981kw} for the nucleon vacuum mass. Approximated forms incorporating only lowest dimension condensates are sometimes used as estimates for in-medium nucleon self-energies~\cite{Finelli:2002na,Finelli:2003fk}: $\Sigma_v = 64 \pi^2 \condensate{q^{\dagger} q} / (3 \borel^2) = 0.36 \unit{GeV} n/n_0$ and $\Sigma_s = -M_N -8\pi^2 \condensate{\bar{q} q} / \borel^2 = -0.37 \unit{GeV} n/n_0$ at $\borel^2 = 1 \unit{GeV^2}$.

Although to be confirmed by dedicated sum rule analysis, it is instructive to understand the impact of four-quark condensates at finite density from naive decoupled self-energy equations linearized in density. For fixed Borel mass $\borel^2 = 1 \unit{GeV^2}$, threshold $s_0 = 2.5 \unit{GeV^2}$ and condensates listed below, the self-energies become independent when a constant $E_- = -M_N$ is assumed; with $\vackappa{q}$ adjusted to yield the vacuum nucleon mass the self-energies are estimated as
\begin{equation}
\label{eq:estimate_sigmav}
\Sigma_v = (0.16 + 1.22 \tmedkappa{v} ) \unit{GeV} \dfrac{n}{n_0} \, ,
\end{equation}
\begin{equation}
\label{eq:estimate_sigmas}
\Sigma_s = - (0.32 + 0.11 \medkappa{s} - 0.31 \medkappa{q} )\unit{GeV} \dfrac{n}{n_0} \, .
\end{equation}
Indeed at small values of $k_F$ the impact of $\medkappa{s}$, $\medkappa{q}$ and $\tmedkappa{v}$ is as follows:
The vector self-energy $\Sigma_v$ only depends on $\tmedkappa{v}$, the scalar self-energy $\Sigma_s$ is effected by $\medkappa{s}$ and $\medkappa{q}$, whereby a negative $\medkappa{s}$ works equivalent to a positive value for $\medkappa{q}$ and vice versa. Comparable effects in $\Sigma_s$ point out that a characteristic value of $\medkappa{s}$ is three times the corresponding absolute value of $\medkappa{q}$.
Whereas this qualitative estimate from Eqs.~\eqref{eq:estimate_sigmav} and~\eqref{eq:estimate_sigmas} is in line with the numerical analysis below for small densities $n<0.7n_0$ corresponding to Fermi momenta $k_F = (3 \pi^2 n / 2)^{1/3}<1.2\unit{fm^{-1}}$, the limit of constant four-quark condensates deviates from the widely excepted picture of cancelling vector and scalar self-energies which can be traced back to competing effects of higher order condensates. Since even in the small density limit for constant four-quark condensates the estimated ratio $\Sigma_v / \Sigma_s \sim \tfrac{1}{2}$ cannot be confirmed numerically, these estimates cannot substitute a numerical sum rule evaluation.

\subsection{Numerical Analysis}
\label{sec:numericalanalysis}
In order to investigate numerically the importance of the three combinations of four-quark condensates entering the sum rule equations~\eqref{eq:sumRuleEquation_s}-\eqref{eq:sumRuleEquation_v} at finite baryon density we perform an evaluation for fixed continuum threshold parameter $s_0 = 2.5 \unit{GeV^2}$ in a fixed Borel window $\borel^2 = 0.8 \ldots 1.4 \unit{GeV^2}$.
Since we are especially interested in medium modifications we use all sum rule equations although chiral-odd sum rule equations have been identified more reliable in the vacuum case~\cite{Jin:1997pb} (however note that instanton contributions might change the relevance of particular sum rule equations~\cite{Dorokhov:1989zw,Forkel:1993hj}).
From Eqs.~\eqref{eq:sumRuleEquation_s}-\eqref{eq:sumRuleEquation_v} after transformation one unique left-hand side and the corresponding three r.h.sides are compared and their differences are minimized by a search for the optimum parameters $\Sigma_v$, $M_N^*$ and $\lambda_N^{*2}$ with a logarithmic deviation measure~\cite{Leinweber:1989hh,Jin:1993up}.
The condensates are estimated from various relations, e.g., the chiral condensate \condensate{\bar{q} q} depends via partial conservation of the axial current (PCAC) on the pion decay constant and pion mass, the gluon condensate $\condensate{ (\alpha_s/\pi) G^2 }$ is determined from the QCD trace anomaly and further condensates can be expressed through moments of parton distribution functions. We use the values
$\condensate{ \bar{q} q }  = - (0.245 \unit{GeV})^3 + n \sigma_N/(2 m_q)$ with
$\sigma_N = 45 \unit{MeV}$ and $m_q = 5.5 \unit{MeV}$,
$\condensate{g_s \bar{q} \sigma G q}  = x^2 \condensate{ \bar{q} q } + 3.0 \unit{GeV}^2 n$ with $x^2=0.8 \unit{GeV}^2$,
$\condensate{\bar{q} iD_0 iD_0 q} + \condensate{g_s \bar{q} \sigma G q}/8  = 0.3 \unit{GeV}^2 n$,
$\condensate{ (\alpha_s/\pi) G^2 }  = -2 \langle (\alpha_s/\pi) (\vec{E}^2-\vec{B}^2)\rangle = -2 [-0.5 (0.33 \unit{GeV})^4 + 0.325 \unit{GeV} n)]$,
$\langle (\alpha_s/\pi) [(vG)^2 + (v\tilde{G})^2] \rangle  = - \langle (\alpha_s/\pi) (\vec{E}^2+\vec{B}^2)\rangle = 0.1 \unit{GeV} n$,
$\condensate{q^\dagger iD_0 iD_0 q} + \condensate{g_s q^\dagger \sigma G q}/12  = (0.176 \unit{GeV})^2 n$,
$\condensate{ q^{\dagger} q }  = 1.5 n$,
$\condensate{ q^{\dagger} iD_0 q }  = 0.18 \unit{GeV} n$ and
$\condensate{ g_s q^\dagger \sigma G q }  = -0.33 \unit{GeV}^2 n$ as employed and discussed in~\cite{Jin:1992id}.
The values of possible $\kappa^{\rm med}$ parameters are given in Eqs.~\eqref{eq:pcqmKappaSet}. Values for $\vackappa{q}$ are adjusted to reproduce the vacuum nucleon pole mass.

\begin{figure}[htb]
\vspace*{-0.1cm}
\includegraphics[width=10.5cm,angle=270]{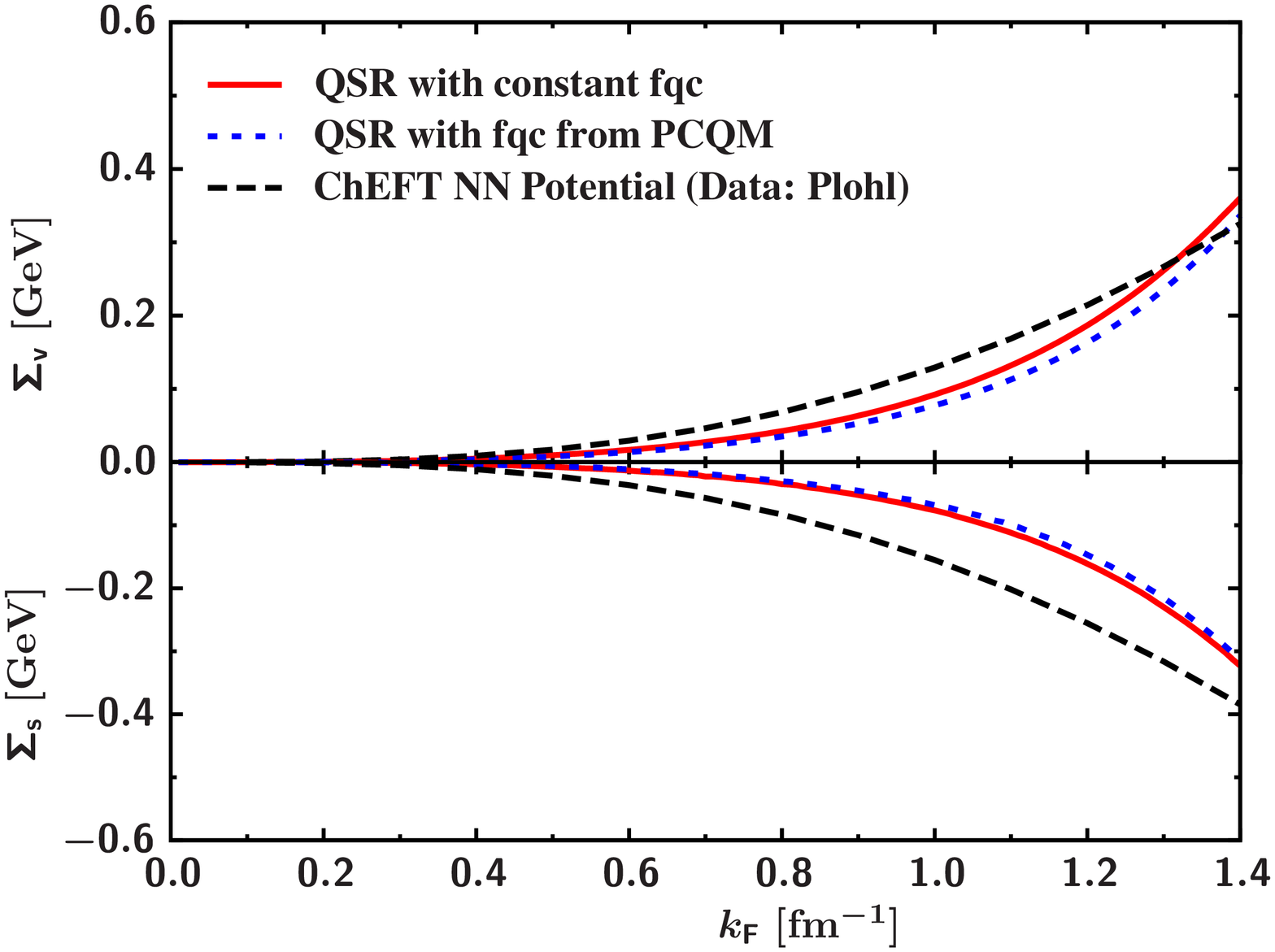}
\vspace*{-0.01cm}
\caption{Nucleon vector and scalar self-energies as functions of the nucleon Fermi momentum $k_F=(3\pi^2n/2)^{1/3}$. The sum rule result for constant four-quark condensates (QSR with constant fqc: $\medkappa{s} = \medkappa{q} = \tmedkappa{v} =0$, solid curve) is compared to an evaluation with density dependent four-quark condensates as given in Eqs.~\eqref{eq:pcqmKappaSet} (QSR with fqc from PCQM, dotted curves). The latter choice causes only minor differences in $\Sigma_v$ and $\Sigma_s$, for the scalar self-energy also because of competing impact of $\medkappa{s}$ and $\medkappa{q}$. The self-energies from chiral effective field theory~\cite{Plohl:2006hy} (ChEFT, dashed curves) are shown as well but should be used as comparison only at small densities.}
\label{fig:basis}
\end{figure}

\begin{figure}[htb]
\vspace*{-0.1cm}
\includegraphics[width=10.5cm,angle=270]{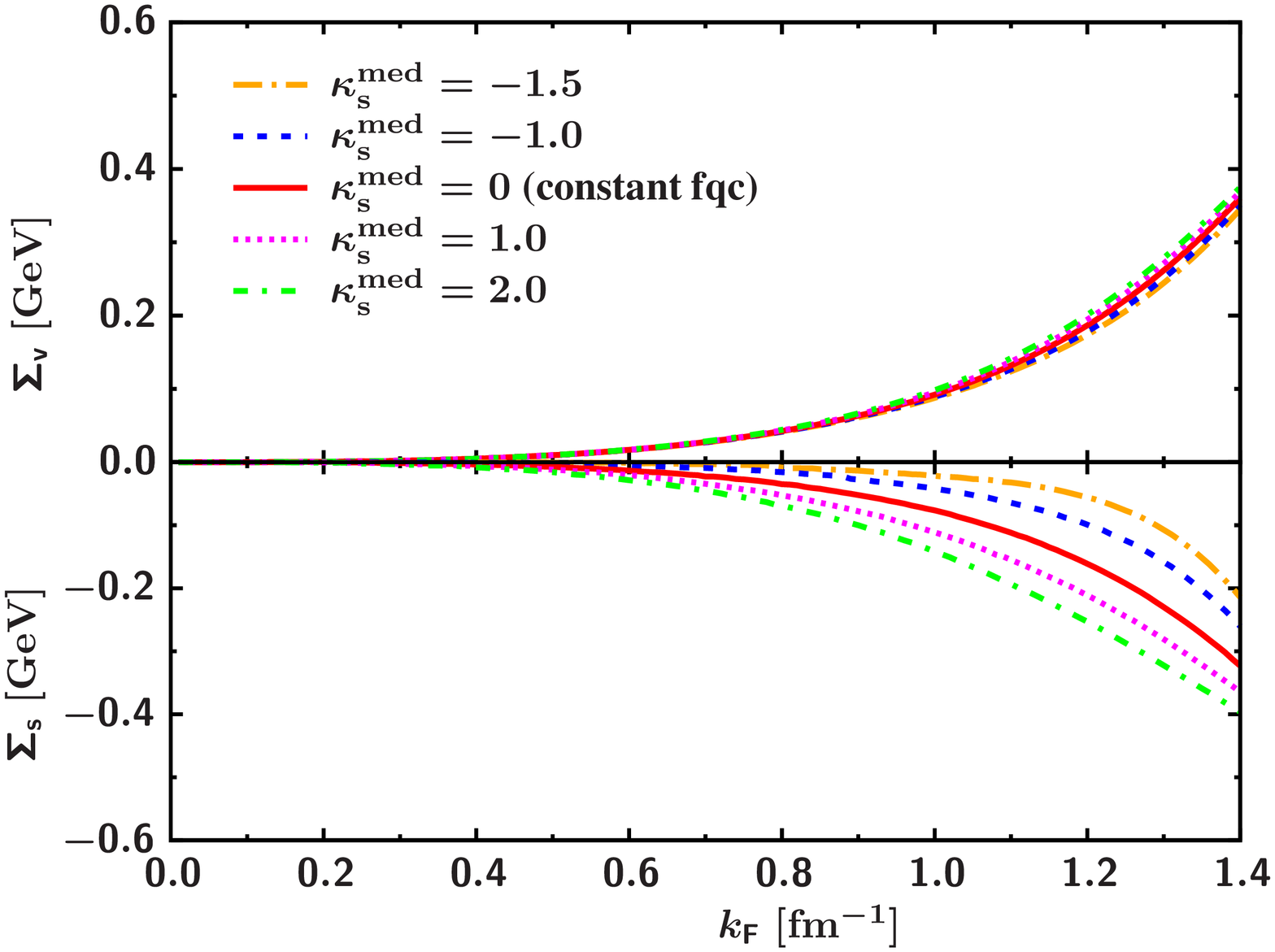}
\vspace*{-0.01cm}
\caption{The variation of nucleon self-energies for different assumptions of the density dependence of the four-quark condensates in Eq.~\eqref{eq:sumRuleEquation_s} parametrized by $\medkappa{s}$; other four-quark condensate combinations are held constant.}
\label{fig:kappa_s}
\end{figure}

\begin{figure}[htb]
\vspace*{-0.1cm}
\includegraphics[width=10.5cm,angle=270]{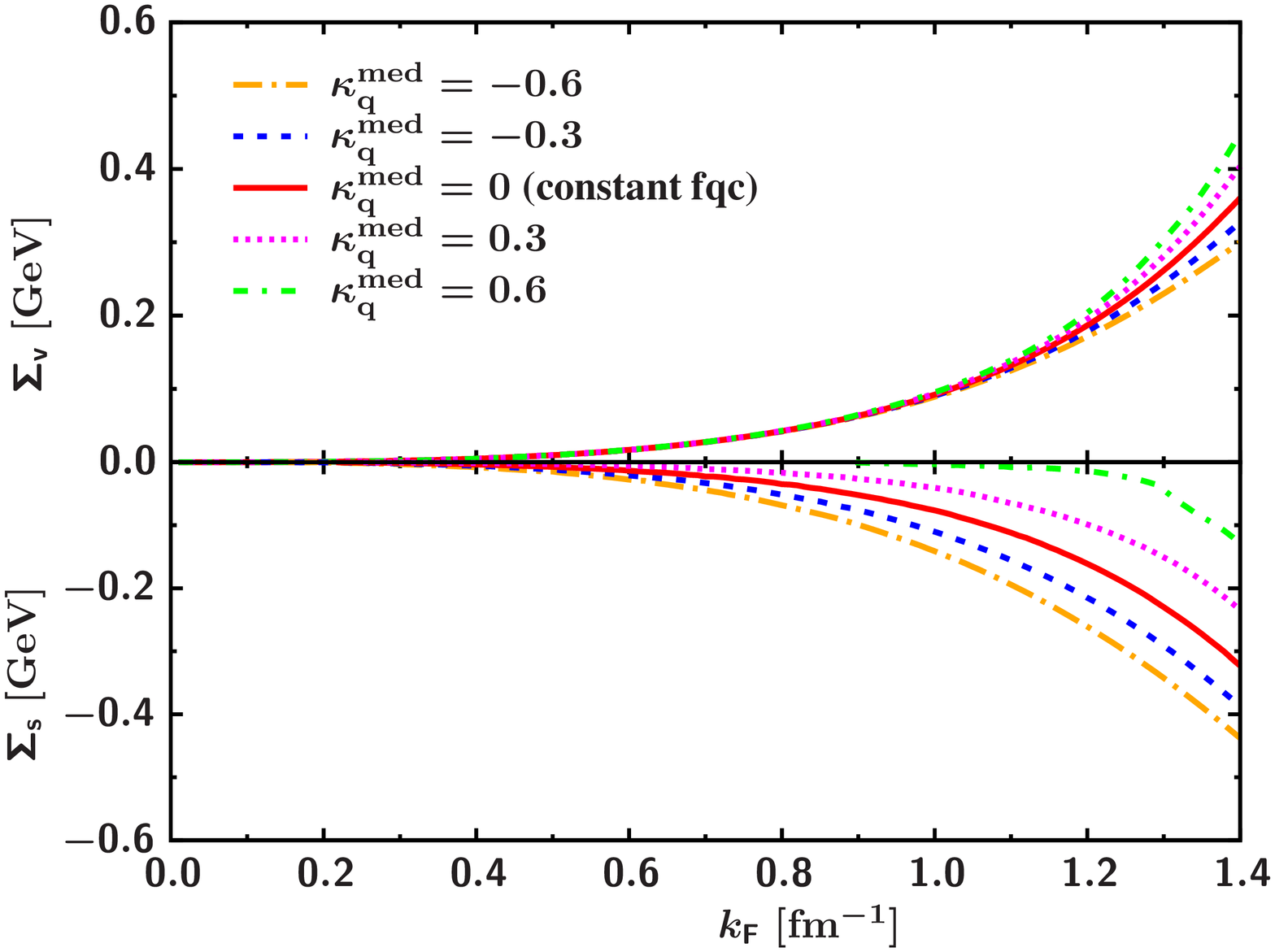}
\vspace*{-0.01cm}
\caption{The same as Fig.~\ref{fig:kappa_s} but for a variation of $\medkappa{q}$ ($\medkappa{s}=\tmedkappa{v}=0$).}
\label{fig:kappa_q}
\end{figure}

\begin{figure}[htb]
\vspace*{-0.1cm}
\includegraphics[width=10.5cm,angle=270]{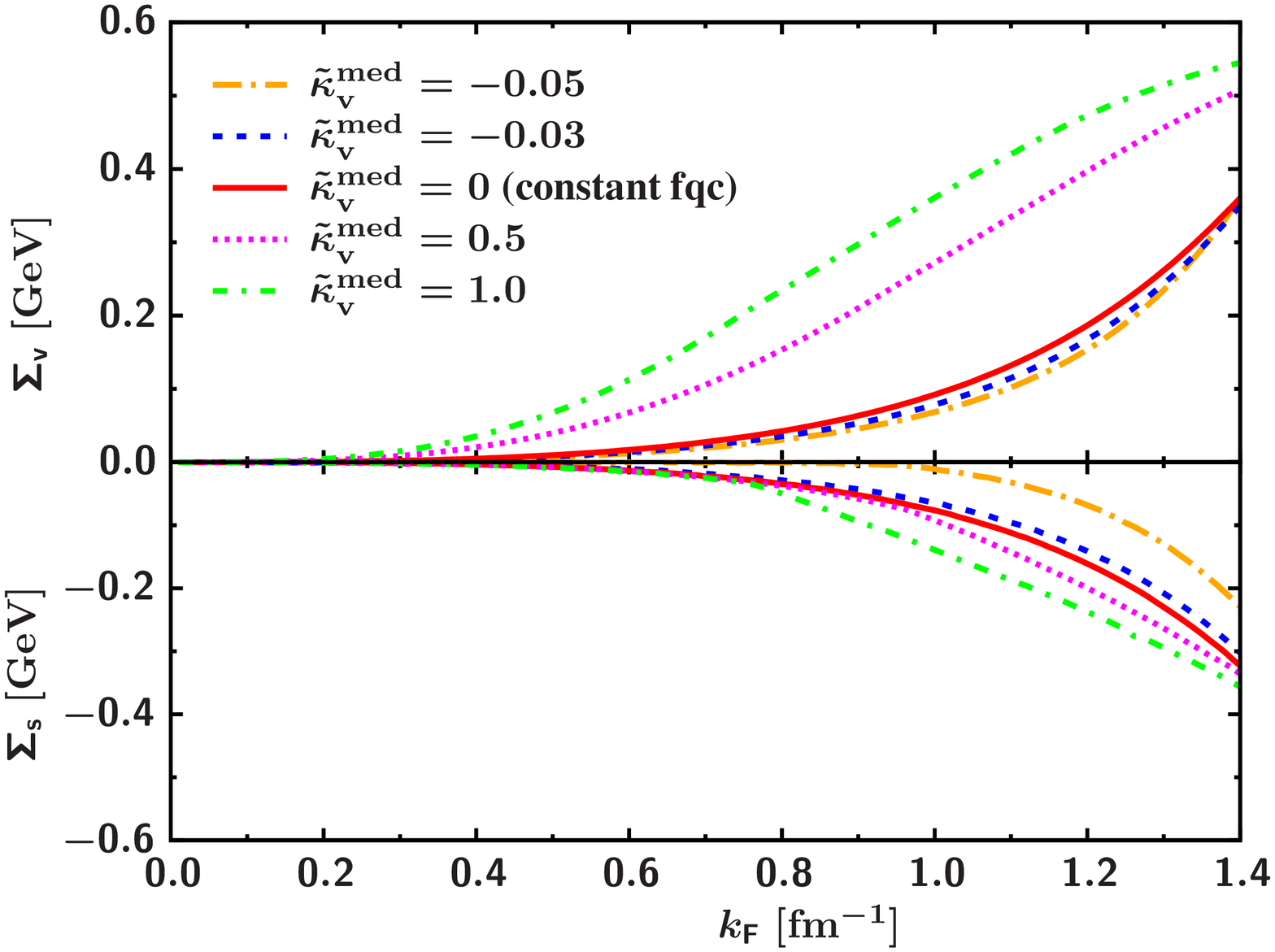}
\vspace*{-0.01cm}
\caption{The same as Fig.~\ref{fig:kappa_s} but for a variation of $\tmedkappa{v}$ ($\medkappa{s}=\medkappa{q}=0$).}
\label{fig:kappa_v}
\end{figure}

\begin{figure}[htb]
\vspace*{-0.1cm}
\includegraphics[width=10.5cm,angle=270]{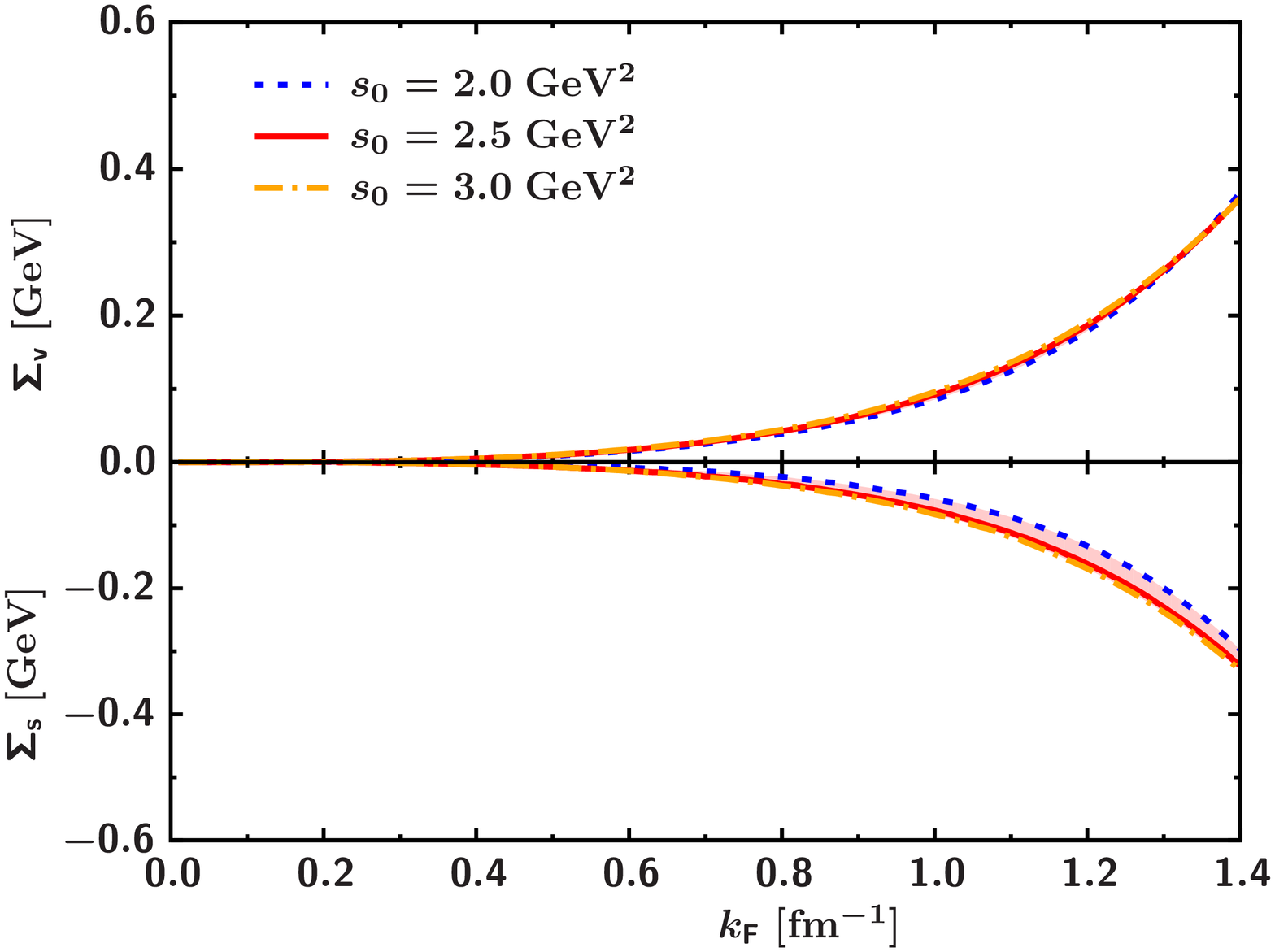}
\vspace*{-0.01cm}
\caption{The impact of different threshold parameters $s_0$ on the nucleon self-energies for the case of constant four-quark condensates, i.\ e. for $\medkappa{s}=\medkappa{q}=\tmedkappa{v}=0$.}
\label{fig:threshold}
\end{figure}

\begin{figure}[htb]
\vspace*{-0.1cm}
\includegraphics[width=10.5cm,angle=270]{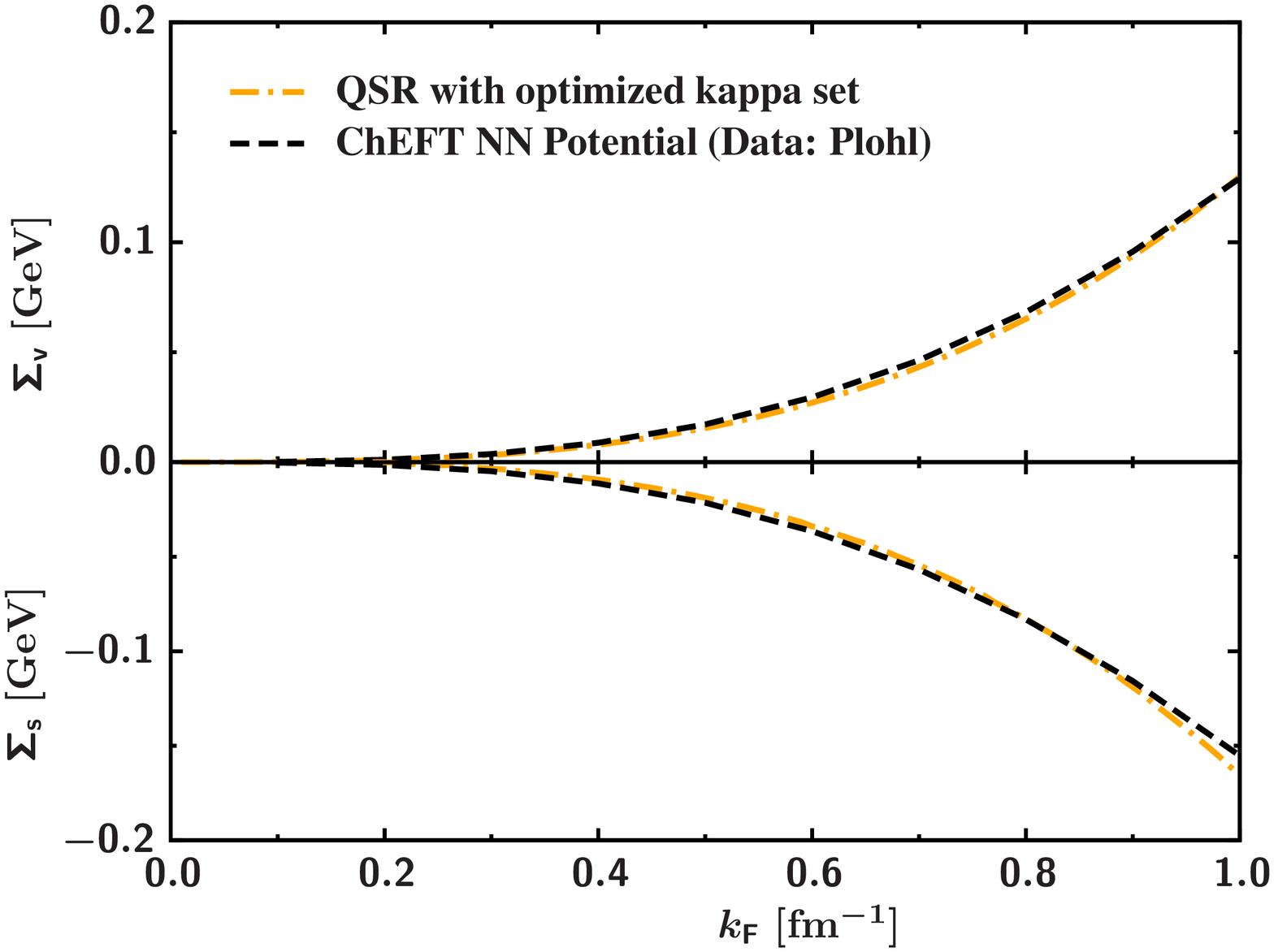}
\vspace*{-0.01cm}
\caption{QCD sum rule evaluations of nucleon self-energies with the parameter set $\medkappa{s}=1.2$, $\medkappa{q}=-0.4$, $\tmedkappa{v}=0.1$ (dash-dotted curves) compare to chiral effective field theory~\cite{Plohl:2006hy} with realistic NN forces as input.}
\label{fig:optimizedkappa}
\end{figure}

The results of these numerical evaluations for a nucleon on the Fermi surface $|\vec{q}_F|=k_F$ are summarized in Figs.~\ref{fig:basis}-\ref{fig:optimizedkappa}.
Fig.~\ref{fig:basis} shows the scalar and vector self-energies of the nucleon as a function of the Fermi momentum. The situation with four-quark condensate combinations~\eqref{eq:fqcparametrization_combination_s}-\eqref{eq:fqcparametrization_combination_v} kept constant at their vacuum value (i.\ e.\ $\medkappa{s}=\medkappa{q}=\tmedkappa{v}=0$) is compared to the QCD sum rule evaluation with $\kappa$ parameters from Eqs.~\eqref{eq:pcqmKappaSet}. The results have the same qualitative behavior as self-energies determined from chiral effective field theory with realistic NN potentials~\cite{Gross-Boelting:1998jg,Plohl:2006hy}. Figs.~\ref{fig:kappa_s},~\ref{fig:kappa_q} and~\ref{fig:kappa_v} exhibit the impact of the 3 different four-quark condensate combinations: The vector self-energy is, in agreement with Eq.~\eqref{eq:estimate_sigmav}, mainly determined by $\medkappa{v}$ especially for small densities (for positive values of $\medkappa{v}$ even the qualitative form of the vector self-energy changes), $\medkappa{q}$ has only small impact, and $\medkappa{s}$ does not effect $\Sigma_{\rm v}$ at all. The scalar self-energy, in contrast, is influenced by all 3 combinations, whereby the change of $\medkappa{v}$ is only visible for Fermi momenta $k_F>0.8 \unit{fm^{-1}}$ as also suggested by Eq.~\eqref{eq:estimate_sigmas}. Figs.~\ref{fig:kappa_s} and~\ref{fig:kappa_q} also reveal the opposed impact of $\medkappa{s}$ versus $\medkappa{q}$. A variation of $s_0$ is not crucial (see Fig.~\ref{fig:threshold}).
The inclusion of anomalous dimension factors in the sum rule equations as in~\cite{Jin:1993up,Cohen:1994wm} leads to a reduction of $\Sigma_v$ in the order of $20 \%$ but causes only minor changes in $\Sigma_s$. Thereby the naive choice of the anomalous dimension from the factorized form of the four-quark condensates leaves space for improvement since it is known that four-quark condensates mix under renormalization~\cite{Jamin:1985su}.
Our analysis concentrates on the impact of four-quark condensates, but also the variation of the density dependence of further condensates can change the result. For example, a large change of the density behavior of the genuine chiral condensate, as determined by the $\sigma_N$ term, by factor 2 \{0.5\} leads to 8~\% decrease \{4~\% increase\} in the effective mass parameter $M_N^*$ at $k_F \sim 0.8 \unit{fm^{-1}}$, while $\Sigma_v$ is less sensitive. Correspondingly, the effective coupling $\lambda_N^{*2}$ is reduced by 10~\% \{enhanced by 5~\%\}.

An improved weakly attractive cancellation pattern between $\Sigma_s$ (attraction) and $\Sigma_v$ (repulsion), and thus agreement with chiral effective field theory~\cite{Plohl:2006hy}, can be achieved for a parameter set $\medkappa{s}=1.2$, $\medkappa{q}=-0.4$, $\medkappa{v}=0.1$ (see Fig.~\ref{fig:optimizedkappa}). However such a fit would allow larger values of $\medkappa{s}$ compensated by a larger magnitude of the negative value of $\medkappa{q}$ and vice versa. Note that in both ways the factorization limit $\medkappa{s,q}=1$ is violated by one or the other four-quark condensate combination. Such optimized $\kappa$ parameters, adjusted to results of~\cite{Plohl:2006hy},  deviate noticeably from those in Eqs.~\eqref{eq:pcqmKappaSet} deduced from~\cite{Drukarev:2003xd}.

Quantities characterizing the energy of an excitation with nucleon quantum numbers are $M_N^*$ and $E_+$, introduced in section 2.4. Since $\Sigma_s$ is negative, $M_N^*$ drops continously with increasing density achieving a value of about $540 \unit{MeV}$ at nuclear saturation density (corresponding to $k_F \sim 1.35 \unit{fm^{-1}}$) if extrapolated from the optimized fit in Fig.~\ref{fig:optimizedkappa}. The energy $E_+$ barely changes as function of $k_F$.

Considering the behavior of the effective coupling parameter in the cases of Figs.~\ref{fig:kappa_s}-\ref{fig:kappa_v} the maximum impact of $\medkappa{s}$ \{$\medkappa{q}$\} on $\lambda_N^{*2}$ is 6~\% \{3~\%\} at $k_F \sim 0.8 \unit{fm^{-1}}$. In the extreme case, $\tmedkappa{v}=1$ leads to a 40~\% increase of $\lambda_N^{*2}$.
The variation of this coupling as a function of $k_F$ is in the order of 10~\% in the optimized scenario. Generally, specific assumptions on the four-quark condensates can cause a decrease or an increase as well. This alternation of $\lambda_N^{*2}$ has already been pointed out in~\cite{Jin:1993up}, whereby their assumptions yield even a $\pm 20~\%$ change at nuclear density compared to the vacuum limit (cf.\ also~\cite{Furnstahl:1995nd}). The vacuum limit of the calculated $\lambda_N^{*2}$ agrees with the existing range of values (see~\cite{Leinweber:1994nm} for a compilation of results for the coupling strength of the nucleon excitation to the interpolating field in vacuum).

\clearpage

%\clearpage

\section{Conclusions}
Four-quark condensates have a surprisingly strong impact on conventional spectral QCD sum rules of light vector mesons. Unfortunately, four-quark condensates and their density dependencies are poorly known. One possibility is to consider a large set of hadronic observables and to try to constrain these parameters characterizing the QCD vacuum. Steps along this line of reasoning have been done, e.\ g.\ , in \cite{Johnson:1995sk}. In order to accomplish a systematic approach, we present here a complete catalog of independent four-quark condensates for equilibrated symmetric or asymmetric nuclear matter. While the number of such condensates is fairly large already in the light quark sector, we point out that only special combinations enter the QCD sum rules. For the conventional nucleon QCD sum rule, three different combinations of four-quark condensates are identified. We note that the knowledge of these combinations (even the individual condensates entering) is not sufficient to convert them into the combination being specific for the spectral QCD sum rule for light vector mesons.
In analyzing the set of independent four-quark condensates we find also identities which must be fulfilled in a consistent treatment. Model calculations of four-quark condensates seem not to fulfill automatically these constraints.

On the level of an exploratory study we show the impact of the three combinations of four-quark condensates on the vector and scalar self-energies of the nucleon. In cold nuclear matter at sufficiently low densities the density dependence of only one effective four-quark condensate combination is found to be important for the vector self-energy and the other two combinations dominate the scalar self-energy. Keeping in mind that the nucleon self-energy pieces per se are not proved to represent observables, one is tempted to try an adjustment of these parameters to advanced nuclear matter calculations. While the overall pattern agrees fairly well (i.\ e.\ large and opposite scalar and vector self-energies) we can reproduce also the fine details on a quantitative level at low densities. Keeping the four-quark condensates frozen in to vacuum values or giving them a density dependence as suggested by a perturbative chiral quark model induce some quantitative modifications which may be considered as estimator of systematic uncertainties related to the four-quark sector. Furthermore, the special use of sum rules and interpolating current and details of the numerical evaluation procedure may prevent QCD sum rules for the nucleon as precision tool. The knowledge of this situation may be of relevance for approaches to the nuclear many-body problem which utilizes chiral dynamics and condensate-related features of the mean field.

Finally, we remind that our study is restricted to cold nuclear matter. The extension towards finite temperature deserves separate investigations.

\section*{Acknowledgement}
Discussions with M. Birse, W. Weise and S. Zschocke are greatfully acknowledged. We thank Ch. Fuchs and O. Plohl for providing extended calculations of nucleon self-energies and S. Leupold for clarifying discussions about four-quark condensate classifications. The work is supported by 06DR136, GSI-FE, EU-I3HP.

\section{Appendix}

\subsection{Operator Product Expansion}
\label{ap:ope}
For completeness and to clarify some technical details we recollect important steps of an OPE calculation. A convenient way to obtain this series is to calculate the Wilson coefficients in
an external weak gluon field \cite{Novikov:1983gd}. In the background field formalism the correlation function~(\ref{eq:correlationFunction}) is expanded
according to Wick's theorem
$\Pi (x) = \Pi_\indexrm{per} (x) + \Pi_\indexrm{2q} (x) + \Pi_\indexrm{4q} (x) + \ldots$ ,
where the full contractions are collected in the perturbative term $\Pi_{\rm per}$ and further terms $\Pi_\indexrm{2q,4q,\ldots}$ denote the number
of non-contracted quark operators. The latter terms give rise to non-local condensates containing the indicated number of
quark operators. The use of Wick's theorem naturally introduces the normal ordering of operators
$\expvaluePsi{:\hat{A}_1 \cdots \hat{A}_n:} \equiv \condensate{\hat{A}_1 \cdots \hat{A}_n}$,
which will be assumed in all expectation values formed out of products of field operators.
%For a short hand notation we will not write out the normal ordering explicitly.

Under the presence of the gluon background field the quark propagator $S^q$ which appears in the terms in $\Pi (x)$ is modified, following from the solution of the Dirac equation in an external field in the Fock-Schwinger gauge for the gluon field.
The corrections to the free quark operator appear in an expansion in the coupling $g_s = \sqrt{4 \pi \alpha_s}$
\begin{equation}
S^q_{ab} (x) = \expvaluePsi{\timeorder{q_a (x) \bar{q}_b (0)}} = \dfrac{i}{2\pi^2} \dfrac{\slash{x}}{x^4} \delta_{ab} + \dfrac{ig_s}{8 \pi^2} \tilde{G}_{\mu \nu}^A (0) T^A_{ab} \dfrac{x^\mu}{x^2} \gamma^\nu \gamma_5 + \ldots \, ,
\label{eq:quarkPropagatorInGluonBackground}
\end{equation}
with the dual gluon field strength tensor $\tilde{G}_{\mu \nu}^A= \tfrac{1}{2} \epsilon_{\mu \nu \kappa \lambda} G^{\kappa \lambda A}$ and color matrices $T^A_{ab}$, valid for massless quarks and inclusion of pure gluon condensates up to mass dimension 4.

The Fock-Schwinger gauge is determined by $(x-x_0)_\mu A^\mu (x) = 0$, and usually one chooses $x_0 = 0$. It allows to express partial derivatives of fields
easily by covariant derivatives which matters when expanding non-local products of such operators.
In general, results are gauge invariant, however technically fixing this gauge has enormous advantages in calculations of Wilson coefficients.  Let us remark, that although the term $\Pi_\indexrm{2q}$
initially contains two uncontracted quark field operators, the expansion of the non-local expectation value into local condensates together with weak gluon fields
resulting from modified quark propagators and the use of the equations of motion would induce further four-quark condensates at the order $\alpha_s$.

The use of the quark propagator~(\ref{eq:quarkPropagatorInGluonBackground}) leads to gluon insertions in the expectation values in $\Pi$ and thus to condensates of higher
mass dimension. To obtain the condensates the expectation values are projected onto all possible Dirac, Lorentz and color scalars obeying symmetry w.r.t.\ time and parity reversal.
This introduces all possible condensates up to the considered dimension,
and having inserted the projections for the specific correlation function offers also the corresponding Wilson coefficients and therefore the OPE \cite{Jin:1992id}.

For example, the non-local diquark expectation value can be projected on color and Dirac structures
\begin{equation}
\condensate{q_{a\alpha} (x) \bar{q}_{b \beta} (0)} = - \dfrac{\delta_{ab}}{12} \sum_{\Gamma} \epsilon_\Gamma \condensate{\bar{q} (0) \Gamma q(x)} \Gamma_{\alpha \beta} \, ,
\label{eq:nonlocalDiquarkExpansion}
\end{equation}
where elements of the Clifford algebra $\Gamma \in \{ \mathbbm{1}, \gamma_\mu , \sigma_{\mu \nu}, i\gamma_5 \gamma_\mu , \gamma_5  \}$, are contracted over Lorentz indices, $\epsilon_\Gamma = \tfrac{1}{2}$ for $\Gamma = \sigma_{\mu \nu}$ and $\epsilon_\Gamma = 1$ otherwise.
A Taylor expansion of the quark operator at $x=0$ in the Fock-Schwinger gauge
\begin{equation}
q(x) = q(0) + x^\mu D_\mu q(0) + \dfrac{1}{2} x^\mu x^\nu D_\mu D_\nu q(0) + \ldots
\end{equation}
leads to additional Lorentz structures, such that the local expansion of the non-local diquark term~(\ref{eq:nonlocalDiquarkExpansion}) up to mass dimension 5 in the expectation values taken at $x=0$ yields
\begin{equation}
\condensate{q_{a\alpha} (x) \bar{q}_{b \beta} (0)} = - \dfrac{\delta_{ab}}{12} \sum_{\Gamma} \epsilon_\Gamma \Gamma_{\alpha \beta} \left ( \condensate{\bar{q} \Gamma q} + x^\mu \condensate{\bar{q} \Gamma D_\mu q} + \dfrac{1}{2} x^\mu x^\nu \condensate{\bar{q} \Gamma D_\mu D_\nu q}
\right ).
\end{equation}
However, matrix elements $\condensate{ \bar{q} (0) \Gamma q(x)}$ with $\Gamma \in \{ \sigma_{\mu \nu}, i \gamma_5 \gamma_\mu, \gamma_5 \}$ do not contribute due to the demand of time and parity reversal invariance and the multiplication with the symmetric Taylor expansion in $x$.
Condensates with field derivatives can be transformed whereby a couple of manipulations using the equations of motion
\begin{equation}
( i \lslash{D} - m ) q = 0 \, , \mspace{50mu}
\bar{q} ( i \overleftarrow{\lslash{D}} + m ) = 0 \, , \mspace{50mu}
D^{AB}_\mu G_B^{\mu \nu} = g_s \sum_f \bar{q} \gamma^\nu T^A q \, ,
\end{equation}
and the representation of the gluon tensor $G_{\mu \nu} = T_A G^A_{\mu \nu}$
\begin{equation}
G_{\mu \nu} = \dfrac{i}{g_s} [D_\mu , D_\nu], \mspace{20mu}
\text{and thus} \mspace{20mu}
\dfrac{1}{2} g_s \sigma G + \lslash{D} \lslash{D} = D^2, \mspace{50mu}
D_{\mu} = \dfrac{1}{2} \left ( \gamma_\mu \lslash{D} + \lslash{D} \gamma_\mu \right ),
\end{equation}
are exploited to yield simplifications in condensate projections. Terms which contain factors of the small quark mass are neglected in these considerations.

Similar projections can be performed for structures which include gluonic parts from the propagator~(\ref{eq:quarkPropagatorInGluonBackground}) and lead to gluon condensates in $\Pi_{\rm per}(x)$ and are also carried out to find the linear combinations of four-quark condensates in $\Pi_{4q}$.
Following this sketched line of manipulations, one arrives at Eqs.~\eqref{eq:evenOddDecomposition}-\eqref{eq:ope-vo}.

\subsection{Alternative Derivation of Pure-Flavor Four-Quark Condensate Interrelations}
\label{ap:constraints}
The constraints between two different color structures of pure-flavor four-quark condensates have been presented in section~\ref{sec:fqcClassification} by analyzing the specific color structure transformation. For the typical baryon
color combination of four-quark condensates the conversion matrix $\hat{B}$~\eqref{eq:fqcMatrixEquation} was derived with the decisive property that it cannot be inverted.
In algebraic terms, the underlying system of linear equations is linearly dependent. This gave rise to the Fierz relations~\eqref{eq:fqcConstraints1}-\eqref{eq:fqcConstraints5}. If one is only interested in these relations, another direct way of derivation exists.
Thereby one considers the ''zero identity''
\begin{equation}
\epsilon^{abc} \epsilon^{a'b'c'} \; \underset{\indexrm{e}}{\bar{q}}^{a'} \underset{\indexrm{f}}{q}^{a} \underset{\indexrm{g}}{\bar{q}}^{b'} \underset{\indexrm{h}}{q}^{b} \; \underset{\indexrm{e,g}}{(\Gamma C)} \; \underset{\indexrm{f,h}}{(C\tilde{\Gamma})} = 0 \mspace{20mu} \text{if} \mspace{20mu} (\Gamma C)^T = -(\Gamma C) \mspace{20mu} \text{or} \mspace{20mu} (C\tilde{\Gamma})^T = -(C\tilde{\Gamma}),
\end{equation}
which can be seen by a rearrangement of the product and renaming of indices (this is the analog discussion as for the choice of possible interpolating fields for the nucleon). Fierz transformation of this relations yields the basic formula
\begin{equation}
\epsilon^{abc} \epsilon^{a'b'c'} \; \bar{q}^{a'} O_m q^a \bar{q}^{b'} O^n q^b \; \trace{\tilde{\Gamma} O_n \Gamma C O^{mT} C} = 0 \, ,
\end{equation}
which gives, with insertion of allowed $\Gamma$ and $\tilde{\Gamma}$, all possible constraints on the color combinations in the sense of the vector $\vec{z}$ in~\eqref{eq:fqcMatrixEquation}. From the non-vanishing possibilities we list only combinations relevant for four-quark condensates and contract them to achieve relations between components of $\vec{z}$:
\begin{equation}
\begin{aligned}
\Gamma = \mathbbm{1}, \tilde{\Gamma} = \mathbbm{1} & \quad \Longrightarrow \mspace{37mu} 0 = -2z_1 + 2z_2 + z_4 + 2z_6 - 2z_8 \, , \\
\Gamma = \gamma_5, \tilde{\Gamma} = \gamma_5 & \quad \Longrightarrow \mspace{37mu} 0 = -2z_1 - 2z_2 + z_4 - 2z_6 - 2z_8 \, , \\
\Gamma = i \gamma_5 \gamma^\alpha, \tilde{\Gamma} = i \gamma_5 \gamma_\beta & \quad \Longrightarrow \quad
\left \{ \begin{aligned}
0 &= - 2 z_1 + z_2 - z_6 + 2 z_8 \, , \\
0 &= - 2z_1 + 2z_2 - 4 z_3 + z_4 - 4 z_5 - 2z_6 + 4 z_7 + 2z_8 \, ,
\end{aligned} \right . \\
\Gamma = i \gamma_5 \gamma^\alpha, \tilde{\Gamma} = \gamma_5 & \quad \Longrightarrow \mspace{37mu} 0 = iz_9 +z_{10} \, .
\end{aligned}
\end{equation}
This set of constraints is equivalent to~\eqref{eq:fqcConstraints1}-\eqref{eq:fqcConstraints5} in section~\ref{sec:fqcClassification}.

\subsection{Four-Quark Expectation Values in the Nucleon}
Supplementary to Tab.~\ref{tab:condensatePartsFromDrukarev} we collect the underlying coefficients to be understood in connection with the work of Drukarev et al.~\cite{Drukarev:2003xd}.

\label{ap:expectationValues}
\begin{table}[h]
\begin{center}
\begin{tabular}{c|cc|cc|c}
\hline
Expectation & \multicolumn{2}{l|}{Parameters in~\cite{Drukarev:2003xd}} & \multicolumn{2}{l|}{Minimal Modification} & {Mean Value} \\
value & $N=p$ & $N=n$ & $N=p$ & $N=n$ & $N=\tfrac{p+n}{2}$ \\
\hline
$U_N^{S,uu}$ & $ 3.94 $ & $ 4.05$    & $ 3.939 $  & $ 4.047 $  & $ 3.993 $ \\
$a_N^{V,uu}$ & $ 0.52 $ & $ 0.51 $   & $ 0.520 $  & $ 0.510 $  & $ 0.515 $ \\
$b_N^{V,uu}$ & $ -0.13 $ & $ -0.02 $ & $ -0.143 $ & $ -0.023 $ & $ -0.083 $ \\
$a_N^{T,uu}$ & $ 0.98 $ & $ 1.02 $   & $ 0.968 $  & $ 1.009 $  & $ 0.989 $ \\
$b_N^{T,uu}$ & $ 0.05 $ & $ < 0.01 $ & $ 0.045 $  & $ 0.007 $  & $ 0.026 $ \\
$a_N^{A,uu}$ & $ -0.45 $ & $ -0.50 $ & $ -0.471 $ & $ -0.502 $ & $ -0.487 $ \\
$b_N^{A,uu}$ & $ -0.06 $ & $ -0.01 $ & $ -0.054 $ & $ -0.009 $ & $ -0.032 $ \\
$U_N^{P,uu}$ & $ 1.91 $ & $ 1.96 $   & $ 2.002 $  & $ 2.030 $  & $ 2.016 $ \\
\hline
\end{tabular}
\end{center}
\caption{Coefficients of pure flavor nucleon four-quark expectation values (in units of $\vaccondensate{\bar{q} q}=(-0.245 \unit{GeV})^3$) as determined in~\cite{Drukarev:2003xd} in the terminology introduced there and modified values from a fine-tuned parameter set which fulfill the constraints~\eqref{eq:fqcConstraints1}-\eqref{eq:fqcConstraints5}. The parameters $\medkappa{s,q}$ and $\tmedkappa{v}$ are finally derived from the right column which shows the result for isospin symmetric baryonic matter.}
\label{tab:parametersDrukarevPureFlavor}
\end{table}
\clearpage
\begin{table}[h]
\begin{center}
\begin{tabular}{c|cc|c}
\hline
Expectation & \multicolumn{2}{l|}{Parameters in~\cite{Drukarev:2003xd}} & {Mean Value} \\
value & $N=p$ & $N=n$ & $N=\tfrac{p+n}{2}$ \\
\hline
$U_N^{S,ud}$ & $ 3.19 $ & $ 3.19$ & $ 3.19 $ \\
$a_N^{V,ud}$ & $ -0.44$ & $ -0.44 $ & $ -0.44 $ \\
$b_N^{V,ud}$ & $ -0.29 $ & $ -0.29 $ & $ -0.29 $ \\
$a_N^{T,ud}$ & $ 0.19 $ & $ 0.19 $ & $ 0.19 $ \\
$b_N^{T,ud}$ & $ 0.18 $ & $ 0.18 $ & $ 0.18 $ \\
$a_N^{A,ud}$ & $ 0.43 $ & $ 0.43 $ & $ 0.43 $ \\
$b_N^{A,ud}$ & $ -0.06 $ & $ -0.06 $ & $ -0.06 $ \\
$U_N^{P,ud}$ & $ -0.20 $ & $ -0.17 $ & $ -0.185 $ \\
$U_N^{VS,ud}$ & $ -0.28 $ & $ -0.21 $ & $ -0.245 $ \\
\hline
\end{tabular}
\end{center}
\caption{As Tab.~\ref{tab:parametersDrukarevPureFlavor} but for coefficients of nucleon four-quark expectation values parametrizing mixed flavor structures as determined in~\cite{Drukarev:2003xd} and the mean values used to calculate medium strength parameters $\medkappa{}$ in isospin symmetric matter. The modifications referring to pure-flavor four-quark condensates are not needed here.}
\label{tab:parametersDrukarevMixedFlavor}
\end{table}
%\clearpage

\end{document}